\documentclass[a4paper,10pt]{article}
\usepackage{RRA4}

\usepackage[latin1]{inputenc}

\usepackage{moreverb}
\usepackage{amssymb,amsmath,amsfonts,theorem}
\usepackage{paralist}
\usepackage{rotating}
\usepackage{epsfig}
\usepackage{multicol}
\usepackage{subfig}
\usepackage{lscape}
\usepackage{subfloat}
\usepackage{color}

\usepackage{amsmath}
\usepackage{amsfonts}
\usepackage{amssymb}

\newtheorem{defi}{Definition}

\usepackage[pdftex]{hyperref}

\def\url{\begingroup \Url}

\RRNo{7230}
\RRdate{December 2009}
\RRauthor{Ochirkhand Erdene-Ochir, Marine Minier, Fabrice Valois and Apostolos Kountouris}
\authorhead{O. Erdene-Ochir, M. Minier, F. Valois and A. Kountouris}
\RRetitle{Resilient networking for wireless sensor networks}
\RRtitle{Protocoles de communication résilients pour les réseaux de capteurs sans fil}
\RRprojet{SWING}
\RRtheme{\THCom}
\titlehead{Resilient networking in wireless sensor networks}
\URRhoneAlpes
\RRmotcle{sécurité, réseaux de capteurs, attaques, survol, résilience.}
\RRkeyword{security, wireless sensor networks, survey, resiliency, attacks.}
\RRabstract{

This report deals with security in wireless sensor networks (WSNs), especially in network layer. Multiple secure routing protocols have been proposed in the literature. However, they often use the cryptography to secure routing functionalities. The cryptography alone is not enough to defend against multiple attacks due to the node compromise. Therefore, we need more algorithmic solutions. 

In this report, we focus on the behavior of routing protocols to determine which properties make them more resilient to attacks. Our aim is to find some answers to the following questions. Are there any existing protocols, not designed initially for security, but which already contain some inherently resilient properties against attacks under which some portion of the network nodes is compromised? If yes, which specific behaviors are making these protocols more resilient? 

We propose in this report an overview of security strategies for WSNs in general, including existing attacks and defensive measures. In this report we focus at the network layer in particular, and an analysis of the behavior of four particular routing protocols is provided to determine their inherent resiliency to insider attacks. The protocols considered are: Dynamic Source Routing (DSR), Gradient-Based Routing (GBR), Greedy Forwarding (GF) and Random Walk Routing (RWR). 
}

\RRresume{

Dans ce rapport nous traitons la sécurité dans les réseaux de capteurs sans fils (WSNs), plus particulièrement à la couche réseau. Dans la littérature, nombreuses propositions tentent de résoudre ces problèmes de sécurité, en utilisant notamment la cryptographie. Cette dernière permet de résoudre les problèmes classiques d'authenticité, de confidentialité et d'intégrité de données. Ainsi, la cryptographie permet d'obtenir une sécurité de base, mais ne permet pas de se prémunir contre les attaques internes quand une partie des noeuds simples sont corrompus. Par conséquent, nous avons besoin des solutions algorithmiques pour compléter.

Notre objectif est de trouver des réponses aux questions suivantes. Existent-ils, les protocoles non conçus initialement pour la sécurité, mais qui contiennent déjà certaines propriétés intrinsèquement résilients contre les attaques internes? Si oui, quels comportements spécifiques les rendent plus résilient?

Ce travail donnera lieu à deux contributions : i) tout d'abord un survol des stratégies de sécurité en général dans WSNs et au couche réseau (modèle OSI) en particulier, puis ii) une étude sur le comportement de protocoles de routage dans les réseaux de capteurs soumis à des attaques réseaux. Les protocoles considérés sont: Dynamic Source Routing (DSR), Gradient-Based Routing (GBR), Greedy Forwarding (GF) et Random Walk Routing (RWR).

}
\begin{document}
\makeRR

\tableofcontents
\newpage

\section{Introduction}\label{intro}

\begin{figure}[ht]
\begin{centering}
        \includegraphics[scale=0.25]{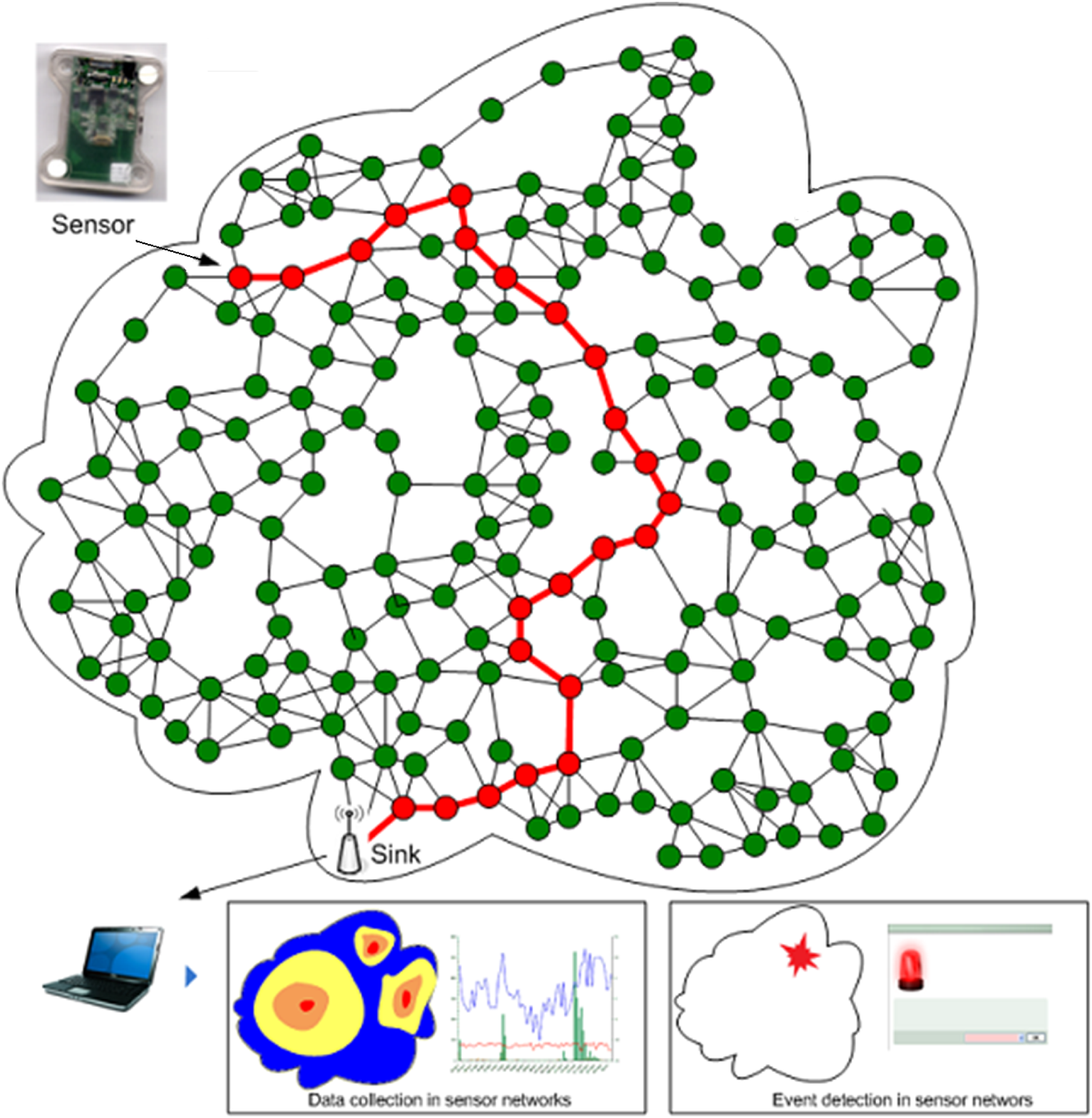} \\
\end{centering}
\caption{Sensor networks} \label{sensornetworks}
\end{figure}

Wireless sensor networks (WSNs) are composed of a large number of low cost and low power, multifunctional sensor nodes (Fig. \ref{sensor}) communicating at short distance. These sensor nodes are densely deployed to collect and transmit data from physical world to one or more destination nodes called ``sink'' in an autonomous way (Fig. \ref{sensornetworks}). WSNs have a wide range of applications such as industrial control, supervising and monitoring, home automation, military applications, detection of environmental parameters, and medical monitoring \cite{Akyildiz2002}. WSNs have multiple advantages such as rapid deployment, cheap, self organized and fault-tolerant.  

WSNs are closer to Ad-hoc networks. They share some common points such as radio communication, decentralized, self-organized and self-configured architecture. Ad-hoc networks are considered to have limited resources, while WSNs have more resource limitations, including a strong energy constraint. Ad-hoc networks utilize point-to-point (``any-to-any'') traffic profile, while WSNs use usually {\it convergecast} (``many to one'') traffic profile. We describe these WSNs characteristics as follow:    

\begin{itemize}
\item Traffic profile
\begin{itemize}
\item Many-to-one: multiple sensor nodes send their sensed data to the sink node. In the presence of several 								 sink nodes, traffic profile is ``many-to-few''. 					
\item One-to-many: the sink node floods control or query information to other sensor nodes. If there are 		                   several sink nodes, traffic is ``few-to-many''.					
\item One-to-any: the sink node can query a specific sensor node.  
\item Any-to-one: a sensor node can send data to the sink node. 
\end{itemize}

\item Hardware constraints
					
\begin{itemize}
\item Limited memory and limited computing power: A sensor is a tiny device with only a small amount of memory and computing power. For example, the common sensor type MICAZ has only 4KB of RAM, 128KB of ROM, 512KB of flash memory and the processor 8 MHz - Atmel AVR Atmega. 

\item Limited power source: A sensor node can only be equipped with a limited battery (<1,5 Ah, 1,2V) \cite{Akyildiz2002}. Because changing batteries in a large number of nodes (of the order of hundreds to thousands) is impractical, sensor node lifetime shows a strong dependency on battery lifetime.

\item Limited power of emission: Because individual sensor nodes have limited radio emission capabilities, accomplishing the network goal often depends on local cooperation by using multi-hop routing. 
\end{itemize}
					
\begin{figure}[ht]
\begin{centering}
        \includegraphics[scale=0.4]{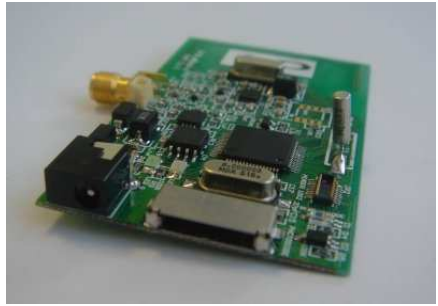} \\
\end{centering}
\caption{Sensor node} \label{sensor}
\end{figure}

\item Environment

Depending on the applications, sensor nodes are frequently deployed in open and hostile geographic areas, unlike a traditional network system in a secure building. They can be deployed in many different ways, from randomly dropping them from a plane or a helicopter, to carefully positioning them one by one \cite{watteyne08energy}. 

\item Communication media

Sensor nodes are linked by a wireless medium, generally, radio communication. Though infrared or optical media \cite{Akyildiz2002} can also be considered as alternatives, both require a line of sight between the sender and receiver, which is impractical for WSNs. 

\item	Topology 

A large number of sensor nodes densely and randomly deployed throughout the sensor field, requires careful handling of topology maintenance. The number of sensor nodes may be of about hundreds or thousands. Depending on the application, the number may reach an extreme value of millions \cite{Akyildiz2002}. The node densities may be as high as 20 nodes/$m^3$ \cite{Akyildiz2002}. The position of sensor nodes is not engineered or predetermined. Topology may change often due to appearance and disappearance of nodes caused by the lack of power, physical damage or environmental interference. Additional sensors can be redeployed at any time. 

\end{itemize}

Security techniques used in traditional network systems cannot be applied directly in WSNs. First, sensors have to be cheap, thus sensors have limited resources (memory, energy, computation power etc.). Second, sensors deployed in an open and hostile environment presenting risks of physical attacks. 

Due to open wireless medium, a passive attacker can eavesdrop communications and an active attacker can alter, replay, and replicate messages. Thus, the confidentiality and the integrity are required. The availability is also required against denial of service (DoS) attacks such as jamming, collisions, exhaustions, etc. Due to unattended devices in open and hostile environments, an adversary can compromise nodes, clone, move nodes, modify software/hardware. Thus, authentication is required.  

The node compromise is the major problem of security in WSNs, which allows an adversary to enter inside the perimeter of security, by extracting sensitive information such as encryption keys, identity, address etc. Subsequently, an adversary can change software/hardware to his own needs: they can produce internal attacks such as Sybil attacks, node replication or Black-Grey-Worm-Sink holes. 

Multiple secure routing protocols have been proposed in the literature \cite{Haas02} \cite{Hu2005ariadne} \cite{Dahill2002} \cite{Perrig2001}. However, even if they often use cryptography to secure routing functionalities, these mechanisms are not relevant against the aforementioned internal attacks stemming from node compromise. Malicious internal nodes can introduce false topological, control, neighborhood, routing information or simply not forward messages. To our knowledge there is not much literature in analyzing routing protocols to determine whether they possess inherent resilient properties to defend against internal attacks, even though they were not initially designed for security. We analyze the resiliency of existing routing protocols in order to emphasize and to augment them by appropriate protocol design focusing on ``beyond cryptography'' approaches. 

\subsection{Report outline} \label{report_outline}
The remainder of this report is organized along the following lines. In \ref{relatedwork} an overview of security issues for WSNs, including existing attacks and defensive measures in \ref{attackdefense}, and security issues of the network layer in \ref{securityNetwork}. Therefore, in \ref{approach}, we explain our approach to the security of routing protocols including a classification of routing protocols and adversary models. Finally, in \ref{simulations}, we present our simulation results and an analysis of behavior of four particular routing protocols to determine their resiliency against attacks. 

\section{Related work}\label{relatedwork}

Compared with other wireless networks such as, for instance, ad-hoc networks and wireless LANs, security in WSNs is more complex because of resource constraints such as limited energy, memory and computational power.  

\subsection{Attacks and defensive measures} \label{attackdefense}
\subsubsection {Ontology } \label{ontology}
According to Wikipedia,

``In computer science and information science, an ontology is a formal representation of a set of concepts within a domain and the relationships between those concepts. It is used to reason about the properties of that domain, and may be used to define the domain.

In theory, an ontology is a ``formal, explicit specification of a shared conceptualization''. An ontology provides a shared vocabulary, which can be used to model a domain - that is, the type of objects and/or concepts that exist, and their properties and relations.''

With this in mind, Znaidi et al. proposed in \cite{Znaidi_Minier_Babau_08} an ontology for attacks in WSNs, composed by four main classes; intention, movement, target and result. 
		
The authors identified five different ``intentions'' such as passive eavesdropping, disruption of 	communication by destroying links, causing unfairness by exhausting available resources (such as bandwidth, energy, battery, etc.), to be authenticated by obtaining access to the network services, and finally, to be authorized by compromising secret information, like for example encryption keys to decrypt messages. 
		
``Movement'' describes the way the attacker reaches one or many of the aforementioned intentions. Technical capabilities of an adversary can be more sophisticated, for example, using a laptop to apply efficient tampering techniques to extract data from sensors. One or many adversary entities can collude to launch a successful attack. Resource constraints or design vulnerabilities in layered network architecture can be exploited by adversaries. 
		
In WSNs, all system resources and network services are potential ``targets'' for the adversaries. A ``target'' can be physical or logical. For example, physical targets can be to destroy the sensor, damage its radio, remove batteries, etc. Examples of logical targets are damaging internal services such as power management, connection between layers, etc., or damaging provided services such as time synchronization, key management, etc.  
		
The authors defined three categories to describe the ``result''. Adversaries can produce only passive damage if an attack can be stopped by some preventive mechanism, partial degradation if a service breaks in one part of the WSN, or broken service for the entire network. 

\subsubsection{Hardware attacks} \label{hardware}

Node compromise is the major problem of security in WSNs. Node compromise allows the adversary to enter inside the perimeter of security. 
		
Because of open and hostile environment deployment, adversaries can easily capture sensor nodes or cause physical damages. Tampering is the well known attack on hardware components of a sensor node that involves modification of its internal structure, allowing an adversary to extract sensitive information such as encryption keys shared between nodes, or even changing the device program to his own needs. Hardware constraints of sensor nodes may facilitate this kind of attacks. However, we should be alert to the technological advancement in the field of tamper-resistance. 
		
The attack via the JTAG interface is described in \cite{Becher2006}. JTAG (Joint Test Action Group) is the interface used for component testing, such as Test Access Port (TAP) and as shown it can allow an adversary to gain full control of a sensor node. They described the attack exploiting the Bootstrap Loader (BSL), where the adversary could read and write on the microcontroller memory. Another simple form of attack is eavesdropping exchanged data on the conductor wires connecting the external memory such as EEPROM and the microcontroller; this enables an adversary to read all transferred valuable data. 
		
Furthermore, adversaries can replicate sensor nodes with the intention to introduce malicious ones. They can also use sensor detection equipment to locate legitimate active sensors and destroy them. 
		
{\bf Defensive measures}
		
Each sensor node can be protected against physical hardware-level attacks by improving its hardware, or by using algorithmic solutions.

Using tamper-resistant hardware, we can protect each sensor node to make sensitive data in its memory inaccessible. In \cite{Becher2006}, the authors propose to disable the JTAG interface or use a good password for the bootstrap loader. Another possible technique is to use special software and hardware outside the sensor to detect physical tampering. Yet an other technique is to use self-termination where a sensor node kills itself, destroys its data and keys when it senses a possible attack. 
		
The algorithmic approach consists in using techniques, such as neighborhood checking, location verification and resilient routing against node compromise. The location based technique consists in making sure the location claims are legitimate. Several researchers have designed routing protocols that achieve some resiliency against node capture by sending every packet along multiple, independent paths and checking at the destination for consistency among the packets that were received \cite{Perrig04secu}. Against attackers with sensor detecting equipment, sensor nodes can work in cooperation by detecting the attacker and prevent other nodes to switch their states \cite{Walters2007}.
		
Chen, et al. proposed to estimate the probability of node compromise in WSNs. The nodes which are close to enemy controlled area may have larger probability to be compromised than the nodes which are far away from enemy controlled area. They describe intelligent models, where a system should have a mechanism to know and record the node compromise events and use current node compromise events to estimate future node compromise accurately.  
		
The following sections summarize existing attacks and defensive measures following the layered network architecture of the OSI model. It is shown that each layer is susceptible to different attacks.

\subsubsection{Physical layer} \label{physical}

The physical layer is responsible for frequency selection, carrier frequency generation, signal detection, 	modulation, and data encryption \cite{Akyildiz2002}.

\begin{itemize}
\item Sniffing, traffic analysis and message corruption. 		\\
		
Because of radio communication nature, an adversary with powerful resources can collect information from sensor nodes if not encrypted. Even if the message transfer is encrypted, they can still acquire enough information to prepare and cause damages. An active attacker can modify the content of message to compromise the integrity. 
				
{\bf Defensive measures}
				
Different cryptographic methods can be used to defend against this kind of attacks.  
		
\item Jamming \\
		
Jamming is a denial-of-service attack, based on the transmission of a radio signal that interferes with the radio frequencies used by the sensor network. Different jamming strategies presented in \cite{Xu2006}:
\begin{itemize}
\item Constant jamming: emitting continuously a radio signal.				
\item Deceptive jamming: instead of sending out random bits, the deceptive jammer constantly injects regular packets to the channel without any gap between successive packet transmissions.
\item Random jamming: instead of continuously sending out a radio signal, a random jammer alternates between sleeping and jamming with the intention of evading detection.
\item Reactive jamming: the jammer will transmit only when it senses channel activity and will stay quiet when the channel is idle.
\end{itemize}
					
Instead of jamming the whole frequency band, an adversary can simply jam the control channel, which is a highly energy efficient and effective jamming strategy.  
			
{\bf Defensive measures} 
			
The standard defense against jamming is using spread-spectrum communication schemes such as frequency hopping spread or direct sequence code spreading. To attack frequency hopping, jammers must be able to follow the precise hopping sequence or jam a wide enough section of the band \cite{Wood2002}. 
			
For the case when attackers jam the broadcast control channel \cite{Chiang07} proposes a defensive measure using a technique based on binary trees. Each emitter constructs a binary tree based on randomly chosen frequencies. The emitter gives to all receivers the frequency corresponding to the leaf of the tree and those of its ancestors. Later the emitter sends two messages simultaneously on two different channels. Jamming is detected when a receiver received the first message and not the second one. 
			
A cryptographic approach can be used \cite{Chan2007} by introducing key assignment to have $t$-resiliency against control channel jamming. Another method proposed in \cite{Tague2009a} is to use random key assignment. If each node has a key, compromising a single node can not affect others, but we require a large number of channels. From this idea they claim that for the key assignment we have to balance the trade-off between 	the number of channels and the robustness against jamming.
			
Other techniques such as channel surfing and spatial retreats are discussed in \cite{Xu2004}. Channel surfing is a form of spectral evasion that involves legitimate wireless devices changing the channel that they are operating on. Spatial retreats are a form of spatial evasion whereby legitimate mobile devices move away from the locality of the jamming emitter. 
			
The technique presented in \cite{Wood03jam} is more of a network layer defense, where sensor nodes may collaboratively map a jammed region and isolate it from the rest of the network. Upon detecting local jamming, nodes blindly report it to their neighbors. Receivers outside the jamming form groups and exchange mapping messages. Groups are coalesced to yield a mapped region.
\end{itemize}

\subsubsection{Link layer} \label{link}

The data link layer is responsible for the multiplexing of data streams, data frame detection, medium access and error control \cite{Akyildiz2002}. The MAC (Media Access Control) layer provides channel arbitration for neighbor to neighbor communication. Cooperative schemes, that rely on carrier sense, let nodes detect if other nodes are transmitting.

\begin{itemize}
\item Collisions \\

Attackers may simply intentionally violate the communication protocol, and continuously transmit messages in order to make interference to generate collisions. To be more effective they can send their own signal when they hear that a legitimate node will transmit. Such collisions would require the retransmission of any packet affected by the collision. In theory, causing collisions in only one byte is enough to create a CRC error and to cripple the message. The advantage of a collision attack compared to a jamming attack is the short power energy consumed and the difficulty to detect it. Corrupted ACK control message could induce costly exponential back-off in some MAC protocols.  
		
{\bf Defensive measures} 
		
All countermeasures that can be used against jamming attack can be applied to collision attacks. Another method is to use error correcting codes \cite{Znaidi_Minier_Babau_08}, which are efficient in situation where errors occur on a limited number of bytes but this solution presents also an expensive communication overhead and additional processing.
		
A method for detecting the MAC layer DoS attacks in a CSMA/CA network is proposed in \cite{Toledo2008}, based on calculating the probability that the collisions in the network can be explained by simple observation of the events in the network. This technique, based on the M-truncated sequential Kolmogorov-Smirnov statistics, monitors the successful transmissions and the collisions of the terminals in the network, and determines how explainable the collisions are given for such observations.

\item Exhaustion \\

Exhaustion attacks consist in introducing collision and force the sensor node to retransmit continuously until exhaustion of battery resources. In \cite{Rachedi2008}, the attacks in IEEE 802.11 are described, which consist in violating of the algorithm BEB (Binary Exponential Backoff), manipulation of parameters such as SIFS, DIFS, EIFS and manipulation of control packets such as RTS (Ready-To-Send), CTS (Clear-To-Send) in order to disturb policy of access to the channel. The adversary can repetitively request channel access with RTS, eliciting a CTS response from targeted neighbor until exhaustion of its resources (virtual jamming). 
 
{\bf Defensive measures} 
		
Most solutions proposed in the literature are trying to reduce the impact of this attack and not to eliminate definitely. One possible defensive measure is to limit the MAC admission control rate and so the sensor network can ignore excessive requests from adversary and prevent energy loss. Another technique is to allow for each sensor node a small slot of time to access to the channel and transmit data, so it limits the possibility of long use of the MAC channel. 

\item Link layer jamming \\

A link-layer jamming attack is presented in \cite{Law2009}. The adversary tries to find a data packet and jam it. However, as packets are generated spontaneously it becomes increasingly difficult for an adversary to predict data packet arrival times. To resolve this difficulty the adversary can look at the probability distribution of the inter-arrival times between all types of packets. This attack can be applied to different MAC protocols such as S-MAC, B-MAC and L-MAC. 
		
{\bf Defensive measures} 
		
Some defensive measures against link-layer jamming are discussed in \cite{Law2009}. In the case of S-MAC, the solution is to prevent clustering based analysis from being feasible by narrowing the distance between the two clusters. In the case of L-MAC, a partial solution is to make the estimation of the clusters more difficult by using pseudo-random function of time to change the slot sizes of packet. For example, a sensor node can change its packet slot size every second by picking a random value from a range. For the B-MAC, the solution is to shorten the preamble in order to make its detection harder. 
\end{itemize}

\subsubsection{Routing layer} \label{routing}

Network layer is responsible for addressing, neighborhood discovering and routing. As WSNs are envisaged to use multi-hop communication, messages may traverse many hops before reaching their destination. The attacks on routing layer are summarized in \cite{Karlof2003}.

\begin{itemize}
\item Sybil attacks \\

Newsome et al. describe in \cite{Newsome2004} the Sybil attack in the context of WSNs. The Sybil attack is defined as a malicious device illegitimately taking on multiple identities (Fig. \ref{sybil}). The malicious nodes can fill their neighbors' memories with non existing neighbors. The Sybil attack is also effective against routing algorithms, data aggregation, and resource allocation. 

{\bf Defensive measures} 

To defend against Sybil attack the network needs some mechanism to validate that a particular identity is the only identity being held by a given physical node. In \cite{Newsome2004} two methods to validate identities are described for direct and indirect validation. In direct validation a trusted node directly tests whether the joining identity is valid. In indirect validation, another trusted node is allowed to vouch for or against the validity of a joining node. Newsome et al. primarily describe direct validation techniques, including a radio resource test. In the radio test, a node assigns each of its neighbors a different channel on which to communicate. The node then randomly chooses a channel and listens. If the node detects a transmission on the channel it is assumed that the node transmitting on the channel is a physical node. Similarly, if the node does not detect a transmission on the specified channel, the node assumes that the identity assigned to the channel is not a physical identity.  

Another technique to defend against the Sybil attack discussed in \cite{Walters2007} is to use random key pre-distribution techniques. The idea behind this technique is that with a limited number of keys, a node that randomly generates identities will not possess enough keys to take on multiple identities and thus will be unable to exchange messages on the network due to the fact that the invalid identity will be unable to encrypt or decrypt messages.

\begin{figure}[ht]
\begin{centering}
        \includegraphics[scale=0.6]{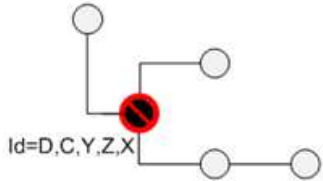} \\
\end{centering}
\caption{Sybil attack.} \label{sybil}
\end{figure}

\item Node replication \\

An adversary may capture nodes, analyze and replicate them, and insert these replicas at strategic locations in the network (Fig. \ref{replication}). Such attacks may have severe consequences; they may allow the adversary to corrupt network data or even disconnect significant parts of the network \cite{Parno2005}. 

{\bf Defensive measures} 

Defenses against node replication attacks can be distinguished into centralized and distributed approaches. In a centralized approach each node will send to a base station a list of its neighbors together with a location. The base station verifies that no node is in two locations at the same time. The two main inconveniences is the existence of a single point of failure (i.e. the base station) and the need to have a permanently present base station.    

In the distributed approach, instead of using a central base station, we could rely on a node's neighbors to perform replication detection. Using a voting mechanism, the neighbors can reach a consensus on the legitimacy of a given node. In node-to-network broadcast approach, each node broadcasts to the whole network a signed location claim and stores the location claims of its $d$ neighbors. If it receives a signed location claim conflicting with one of its neighbors it broadcasts to the whole network a revocation proof containing the conflicting claims. If $n$ represents the number of nodes and $d$ the number of witnesses, in this case the communication cost is $O(n)$  and memory usage is $O(d)$.

In a deterministic multicast approach, there is a public deterministic function {\bf F} that for each node $i$ outputs a set of witness nodes $F(i)$ \cite{Aguilar2009}. Communication cost is $O(d X lnd X n X \sqrt{n})$ and memory usage is $O(d)$.

In \cite{Parno2005} the authors propose two distributed algorithms: randomized multicast and line-selected multicast. Randomized multicast improves upon the security of deterministic multicast by randomly choosing the witnesses. The birthday paradox  suggests that there will be at least one collision. Communication cost is $O(n)$ and memory usage is $O(\sqrt{n})$.

The line-selected multicast algorithm seeks to reduce the communication costs by choosing as witnesses intermediate nodes between source and destination. In the protocols proposed in \cite{Parno2005} the nodes detect replication passively. If there is a replicated node, a witness will (passively) receive two conflicting locations claims and use them to ban the replicas. Communication cost is $O(\sqrt{n})$ and memory usage is $O(\sqrt{n})$.

A novel active approach is proposed in \cite{Aguilar2009}. The idea is that each node will actively test if $d$ other random nodes are replicated or not (called scrutinized nodes). The active approach needs a constant number of scrutinized nodes per node so memory usage per node is $O(1)$ and communication costs $O(\sqrt{n})$. 
	
\begin{figure}[ht]
\begin{centering}
        \includegraphics[scale=0.6]{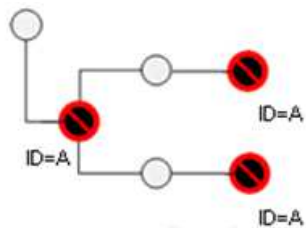} \\
\end{centering}
\caption{Node replication attack.} \label{replication}
\end{figure}

\item Selective forwarding \\

In multi-hop routing, messages may traverse many hops before reaching their destination. In a selective forwarding attack, malicious nodes simply drop some of the messages instead of forwarding all of them (Fig. \ref{selective}). One example of such attack is the black hole attack where the attacker decides to drop all messages. To increase effectiveness of such attacks, attackers will try to put malicious nodes close to base stations to attract more traffic. 

{\bf Defensive measures} 

One possible solution is to use traffic monitoring to ensure that  neighbor nodes forward the messages. In \cite{Marti2000}, the authors propose to use a Watchdog scheme that identifies selfish nodes and a Pathrater scheme that helps routing protocols avoid such nodes. The Watchdog scheme is further extended by a Reputation based scheme, \cite{Michiardi2002}, where the neighbors of any single node collectively rate the node according to how well the node executes the functions requested upon it. 

Another possible way is to use of multi-path routing \cite{Ganesan2001}. These defenses may decrease the probability that a message will encounter an adversary along all routes. In \cite{Roosta2007} the authors analyze single path and multipath routing where deterministic and probabilistic selection mechanisms are used. They show that multi-path routing protocols have better end-to-end packet delivery than single path routing. However, as expected multi-path routing consumes much more energy than the single path routing.

\begin{figure}[ht]
\begin{centering}
        \includegraphics[scale=0.6]{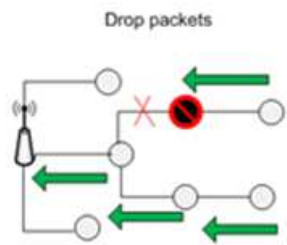} \\
\end{centering}
\caption{Selective forwarding attack.} \label{selective}
\end{figure}

\item Sinkhole attacks \\

In sinkhole attacks, an adversary attracts the traffic to a compromised node. To create a sinkhole the attacker can place a malicious node closer to the sink to attract most of the traffic (Fig. \ref{sinkhole}). After successful sinkhole attacks, adversaries can employ selective forwarding. The nature of sensor networks where all the traffic flows towards one (or few) sink node makes this type of attacks highly relevant.

{\bf Defensive measures} 

One approach to avoid sinkholes is to use routing protocols \cite{Karlof2003} that verify the bidirectional reliability of a route with end-to-end acknowledgments which contain latency and quality information.

\begin{figure}[ht]
\begin{centering}
        \includegraphics[scale=0.6]{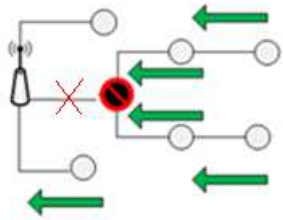} \\
\end{centering}
\caption{Sinkhole attack.} \label{sinkhole}
\end{figure}

\item Wormhole attacks \\

In a wormhole attack as defined in \cite{Karlof2003}, an attacker receives packets at one point in the network and tunnels them to another point in the network via an out-of-band connection (Fig. \ref{sinkhole}). The wormhole attack is particularly dangerous against routing protocols. An adversary situated close to the sink node may be able to completely disrupt routing by creating a well placed wormhole. 

{\bf Defensive measures} 

Wormhole attacks are very difficult to detect, especially when combined with a Sinkhole attack. A wormhole detection mechanism, called packet leashes, is introduced in \cite{Hu2003a} and is based on distance estimation; it consists in two mechanisms: geographical leashes and temporal leashes. The main idea of both mechanisms is to add some information to the packets that restricts their maximum allowed transmission distance. A geographical leash ensures that the recipient of the packet is within a certain distance from the sender. A temporal leash ensures that the packet has an upper bound on its lifetime, which restricts the maximum travel distance, since the packet can travel at most at the speed of light. Either type of leash can prevent the wormhole attack, because it allows the receiver of a packet to detect if the packet traveled further than the leash allows. However, both geographical and temporal leashes require authentication and integrity otherwise an adversary can modify or forge them altering distance information. The main disadvantage of geographical leash is that requires the nodes to be equipped with GPS or to be able to determine their location in some other way. The main disadvantage of time leash mechanism is that it requires extremely tight time synchronization, which might not possible to achieve in some environment. 

Another technique to defend against wormhole attacks consists in using directional antennas \cite{Hu2004}. The main idea is that when two nodes are within each other's communication range, they must hear each other's transmission from opposite directions.  Directional antennas are less expensive than many mechanisms proposed for localization including more efficient use of energy and better spatial use of bandwidth. However, use of directional antennas cannot be afforded in many applications. Another disadvantage is that a link can be lost as the probability of losing links depends on the density of the network. Moreover, this method allows detecting only a single wormhole attack. 

In \cite{Znaidi2008}, the authors use neighborhood information to detect wormhole attacks. The main idea is that if two nodes are declared as 1-hop neighbors and if the network is sufficiently dense, these two particular nodes must have some common 1-hop neighbors whereas it is not the case if the corresponding link is a wormhole.

\begin{figure}[ht]
\begin{centering}
        \includegraphics[scale=0.6]{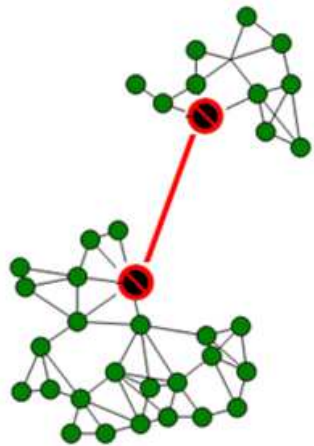} \\
\end{centering}
\caption{Wormhole attack.} \label{wormhole}
\end{figure}

\item Hello flood attacks \\

Many routing algorithms use hello packets for neighborhood discovery. In the hello flood attack described in \cite{Karlof2003}, the attacker tries to convince all nodes to choose it as a parent using a powerful radio transmitter to bomb the whole network with hello messages announcing false neighbor status. So legitimate nodes will attempt transmission to the attacking node despite many of them being out of range. 

{\bf Defensive measures} 

If the attacker has the same reception capabilities, one way to avoid the hello flood attacks is to verify the bi-directionality of local links \cite{Karlof2003}. If not, authentication is a solution for nodes to verify the identity of their neighbors.  

\end{itemize}

\subsubsection{Application layer} \label{application}

\begin{itemize}
\item {Data aggregation}\\

Data aggregation can greatly help to reduce energy consumption by eliminating redundant data in WSNs. The aggregator nodes collect data using suitable aggregation functions \cite{Wagner2004} and then transmit the aggregated result to an upper aggregator or to the sink node. 

The aggregators are vulnerable to attacks especially they are not equipped with tamper resistant hardware. Once a single node is compromised, it is easy for an adversary to inject false data into the network and mislead the aggregator to accept false readings. 

{\bf Defensive measures} 

Numerous techniques proposed in the literature to secure data aggregation. This techniques summarized in \cite{Walters2007}, \cite{Alzaid2008} and more recently \cite{ChenMakki2009}. 

Against node compromise, in \cite{HuEvans03} the authors propose a method where sensors measurements are forwarded unchanged and then aggregated at the second hop instead of aggregating at the immediate next hop. 

This method is improved in \cite{Jadia2004} by using one-hop and two-hop pairwise key instead of $µ$TESLA. A secure hop by hop data aggregation is proposed in \cite{Yang2008} that can tolerate more than one compromised node.

A stealthy attack is discussed in the literature, where an adversary provides false aggregation results without revealing its presence. A framework called {\it aggregate-commit-prove} is proposed in \cite{Przydatek2007}. This framework is extended in \cite{Chan2006} to fully distributed network model instead of the single aggregator model. 

A mathematical framework based on statistical estimation theory and robust statistics to quantify the resiliency of aggregation functions against malicious data is presented in \cite{Wagner2004}. The paper claims that some aggregation operators such as average, sum, minimum and maximum are inherently insecure. They showed that the median is a more robust alternative for averaging data. The authors argued that trimming and truncation can be used to strengthen the security of many aggregation primitives by eliminating possible outliers, thus, providing inherent robustness against attacks.  However, this approach ignores the structure of the network and focus only on the aggregation function at the base station. Moreover, this method can falsify some results by elimination of outliers for some applications such as monitoring bush-fire where outliers carry the useful information. A statistical en-route filtering mechanism is presented in \cite{Ye2005} to detect and drop false reports during the forwarding process. 

Other methods such as a witness based data aggregation, proposed in \cite{Du2003}, assures the validation of the data sent. In \cite{Mahimkar2004} improved data integrity is achieved by signing the aggregated data. A dynamic en-route-filtering scheme for false data injection attacks is proposed in \cite{Yu2005}. Sanli, et al. (2004) developed a new approach, called secure reference based data aggregation, in which sensors send only the difference between sensed data and reference values instead of the raw data.  

During the last few years {\it homomorphic} encryption schemes have been studied extensively. An homomorphic cryptosystem is a cryptosystem that allows direct computation on encrypted data by using an efficient system \cite{Alzaid2008}. It can be used in secure aggregation to provide the end to end privacy needed. The protocol called concealed data aggregation is proposed in \cite{Girao2006} and uses an additive and multiplicative homomorphic encryption scheme. Their work, based on the privacy homomorphism (PH) proposed in \cite{domingo02}, allows direct computation on encrypted data. The authors argue that the security level is still reasonable and PH helps to implement encryption in WSN, although Wagner (2003) proved that PH is insecure against chosen plain text attacks.  A new secure data based on homomorphic encryption is proposed in \cite{Castelluccia2005}. It uses a modular addition instead of the {\it xor} operation that is found in stream ciphers. Thus, even if an aggregator is being compromised, original messages can not be obtained by an attacker. 
\end{itemize}

\subsection{Network layer security} \label{securityNetwork}

An adversary can interfere with the network layer in two ways; from the outside and from the inside of the network. We can state numerous external attacks such as eavesdropping, injecting replayed or fabricated messages. The basic security requirements such as integrity, confidentiality, non-repudiation and authenticity can be ensured by usual cryptographic protections. 

\subsubsection{Security associations (Key management)} \label{management}

Cryptography is the basic encryption method used in implementing security. In WSNs, the asymmetric cryptography is considered expensive for individual nodes in terms of computing power, memory used and consumption of energy. Asymmetric cryptography consists in maintaining two keys one of which is made public and the other is kept private. The main techniques used to implement asymmetric cryptosystems are RSA and Elliptic Curve Cryptography (ECC). For the same level of protection key length used in ECC can be as small as 163 bits rather than the 1024 bits required in RSA \cite{Walters2007}.

Symmetric cryptography is considered as a typical choice for WSN applications. It uses a single shared key known only between communicating nodes. This shared key is used for both encryption and decryption of messages. 

Some approaches adopt both asymmetric and symmetric cryptographic schemes to reduce the overheads. An ``end-middle-end'' security framework is proposed in \cite{Mache2008}, and consists in using a lightweight asymmetric cryptography scheme. They exploit the heterogeneity of sensors by introducing rich gateway nodes with public key cryptography to compute digital signatures. Therefore, the regular nodes can use a symmetric cryptography until a gateway is reached. 

The major symmetric cryptography difficulties in the presence of compromised sensor nodes are key distribution and tamper resistance.  Key distribution protocols can be distinguished into two kinds of approaches; centralized and distributed:

\begin{itemize} 
\item A centralized key management scheme consists in having only one central point called KDC (Key Distribution Center) to ensure creation and distribution of keys. 

\item Distributed key management schemes do not have a centralized entity for creation and distribution of keys. In this approach we can list different methods such as deterministic, probabilistic, localization and $t$-secure protocols \cite{Znaidi2009}. In $t$-secure protocols an adversary has to compromise at least $t+1$ nodes to find a used key. In \cite{Znaidi2009}, the authors propose to use semi-symmetric, $t$-degree Polynomial property by introducing a new variant. This protocol is $t$-secure and the new variant guaranties authentication of identity and consolidates security. 

Another novel approach discussed in \cite{Azimi-Sadjadi2007}, is to couple the physical layer with key generation algorithms. This coupling is based on the wireless communication phenomenon known as the principle of reciprocity of the wireless channel. Fading graphs can be used to generate cryptographic keys, and the non-stationary characteristics of a wireless channel can be used to extract enough entropy to obtain cryptographically secure keys. Hence, this method is more suitable for indoor applications. It is virtually impossible for a third party, which is not located at one of the transceiver's position, to obtain or predict the exact signal envelop. In this case, the radio communication can be considered as an advantage.
\end{itemize}

An adversary when he gains access to the network becomes even more dangerous. Since nodes are usually not well protected physically, they can be captured and compromised allowing an adversary to extract sensitive information such as encryption keys, identity, address etc. In this case, the major functions of the network layer such as: naming-addressing, neighborhood discovery and routing, become targets of internal attacks such as Sybil, node replication or Black-Grey-Worm-Sink holes as described in \ref{routing}.

\subsubsection{Naming and addressing} \label{naming}

Sensor nodes depend on naming and addressing for the routing protocol to be able to convey traffic to them. A naming or addressing scheme usually requires that each node be assigned a unique name or address, in order to avoid ambiguities. An insider adversary can try to break this principle by cheating on the identities. In Sybil attacks an adversary assigns different identities to the same node whereas in node replication attacks several adversary nodes may share the same identity. 

\subsubsection{Neighborhood discovery} \label{discovery}

In many WSN protocols sensor nodes must be aware of their neighborhood to know other nodes they can communicate with. The neighborhood discovery protocol, sometimes also called ``Hello protocol'', nodes broadcast the ``Hello message'' in order to be discovered by their neighbors. An outsider attacker can perform a jamming attack to prevent two nodes from establishing a neighborhood relationship. In addition, an insider adversary can disrupt the neighborhood discovery by introducing false HELLO packets or by not respecting the timing of periodic sends to disrupt the stability of the neighborhood.

\subsubsection{Secure routing} \label{secureRouting}

WSNs use multi-hop communication to increase network capacity and save overall energy (increasing thus the network lifetime) and to reduce interference between nodes. In multi-hop routing, messages may traverse many hops before reaching their destinations. Simple sensor nodes are usually not well physically protected because they are cheap and can be deployed in open or hostile environments where they can be easily captured and compromised (e.g. by an adversary that can extract sensitive information). After node compromise, an adversary gains access to the network and can produce malicious activities. Such attacks may have severe consequences; they may allow the adversary to corrupt network data or even disconnect significant parts of the network. 

We focus here on attacks that appear in the network after node capture. Other attacks such as jamming, exhaustion, collisions, link layer jamming or attacks against data aggregation can be produced in WSNs, however they target essentially the physical, link or application layers. Attacks at the network layer are summarized in \cite{Karlof2003} as follows: (a) spoofed, altered or replayed routing information; (b) selective forwarding, node replication, Sybil attacks or Black-Grey-Sink holes and HELLO flooding. 

The attacks which are within the scope of the present work correspond to those in the second list. These are general attacks which can be performed in any routing protocol. However, numerous attacks target vulnerabilities due to the behavior of a specific protocol. In \cite{Wood2002}, Wood et al. 
underline the necessity to protect protocols at design time. They explained how an adversary can exploit the protocols vulnerabilities to perform efficient DoS attacks. An adversary can exploit reasonable approaches for power conservation and efficiency, which make protocol behavior deterministic and predictable and thus vulnerable to attacks. 

There are some other attacks against specific routing techniques such as rushing attacks, which target on-demand routing protocols, attacks that disrupt route discovery process, location disclosure attacks which target geographic routing protocols, attacks against virtual coordinates and so on. 

Numerous approaches exist to secure routing protocols. 

\begin{itemize} 
\item Cryptographic approaches \\

Here, we elaborate on techniques based on cryptographic approaches to defend against attacks at the routing layer by introducing special secure routing algorithms. 

SRP, ARIADNE, ARAN are proposed to defend against attacks for on-demand routing. The SRP (Secure Routing Protocol), cite{Haas02}, combats attacks that disrupt the route discovery process and guarantees the acquisition of correct topological information. SRP allows the initiator of a route discovery to detect and discard bogus replies. However, SRP is not immune against wormhole attacks, and colluding malicious nodes can misroute the routing packets. 

ARIADNE \cite{Hu2005ariadne} is a secure ad-hoc routing protocol based on DSR, which guarantees that the target node of route discovery process can authenticate the initiator and the initiator can authenticate each intermediate node on the path to the destination. ARIADNE provides point-to-point authentication of a routing message using message authentication codes (MAC) and shared keys between the two parties. The authentication of broadcast packet such as RREQ, ARIADNE is based on the TESLA broadcast authentication protocol. 

ARAN \cite{Dahill2002} is a secure on demand routing protocol that detects and protects against malicious actions carried out by third parties. ARAN introduces authentication, integrity and non-repudiation by using a trusted certificate server. The goal of ARAN is to allow verifying that an intended destination was reached. However, using asymmetric cryptography makes ARAN very costly in term of energy consumption and computation power, hence not suitable for WSNs.  

SEAD \cite{Perkins1994} is a secure proactive routing protocol based on the DSDV (Destination Sequenced Distance Vector) protocol. The basic idea of SEAD is to authenticate the sequence number and metric of a routing update message using one-way hash chains. Moreover, the receiver of SEAD routing information authenticates the sender, ensuring that the routing information originates from the correct node. Authors propose to use TESLA for broadcast authentication or to use MAC assuming shared keys between each couple of nodes in the network. However, SEAD doesn't cope with wormhole attacks. 

SPINS, proposed in \cite{Perrig2001}, is composed by two protocols $µ$TESLA and SNEP. $µ$TESLA introduces asymmetry through delayed disclosure of symmetric keys resulting in an efficient broadcast authentication scheme adapted in WSNs. SNEP provides data confidentiality, two-party data authentication and data freshness. The main goal of SNEP is to protect communication between base station and sensors or between two sensors in the network. SNEP requires a symmetric key initially shared between nodes and base stations. This shared key allows each sensor to deduct the encryption and authentication keys. SNEP proposes to use a counter shared between nodes and base station to guarantee data freshness. 

In \cite{Tubaishat2004}, nodes are divided into different levels depending on the energy consumption and reliability. Low-level nodes have the role to sense and upper-level nodes are responsible for routing and aggregating data. They propose to use symmetric key based on group key management, where every node contributes its partial key to compute the group key. 
	
\item Reputation based schemes \\

The basic idea of a reputation based scheme is to choose the nodes with good reputation for constructing routing paths. Wachdog and Pathrather\cite{Marti2000} are discussed as a trust-based routing scheme. Watchdog allows identifying misbehaving nodes and Pathrather helps to avoid these nodes. Other methods such as Virtual currency, Nuglets, and Sprite are based on the compensation of good contributing nodes by micro credits. Nodes receive virtual payment for forwarding a message and this payment is deducted from the sender. However, such reputation based scheme are mainly designed for Ad-hoc networks and they don't take into account colluding malicious nodes.

\item Multi-path routing \\

Multi-path schemes provide more reliable routing, though they introduce more communication overhead. Some multi-path routing techniques are discussed in the literature in order to resist against node failures. Examples are the disjoint multipath and braided multi-path techniques \cite{Ganesan2001}. Another technique described in \cite{Kim2005} is to repair broken links by utilizing location information. These techniques can be adapted to secure routing against compromised nodes. \cite{Roosta2007} shows that multi-path routing protocols have better end-to-end packet delivery than single path routing, but as expected they consume much more energy. 
	 
\end{itemize}

\section{Our approach} \label{approach}

\subsection{Motivation} \label{motivation}

Node compromise is the major problem of security in WSNs, since it allows an adversary to enter inside the perimeter of security, by extracting sensitive information such as encryption keys, identities, addresses etc. After node compromise, the attackers can produce internal attacks such as Sybil attacks, node replication or Black-Grey-Worm-Sink holes.

As described in \ref{secureRouting}, numerous approaches exist to secure WSNs routing protocols. Most of the existing mechanisms are based essentially on cryptographic primitives \cite{Haas02,Hu2005ariadne,Dahill2002,Perrig2001}, reputation based schemes \cite{Marti2000,Michiardi2002} or use specialized hardware \cite{Hu2004,Hu2003a}. However, these mechanisms do not protect against all attacks and are most of the time inefficient when node compromise is considered. Furthermore, reputation based schemes are specific to selective forwarding attacks. Finally, solutions that use specialized hardware are specific to Wormhole attacks. 

The literature is rather scarce, \cite{Roosta2007}, in analyzing inherent security of routing protocols which though not initially designed for security, may possess inherent resiliency against internal attacks. \cite{Roosta2007} shows that multi-path routing protocols have better end-to-end packet delivery than single path routing, but they consume much more energy. However, they do not take into account routing functionalities, such as route discovery, route maintaining, neighborhood discovery etc. and they focused only on the multipath aspect. With the aim to contribute in this direction, in the next subsection our definition of resiliency is given.

\subsection{Definition of resiliency} \label{securitymetrics}

According to Webster \cite{resilience}, in mechanics resiliency is the capability of a strained body to recover its size and shape after deformation caused especially by compressive stress. In a broader context Webster also defines resiliency as \textit{an ability to recover from or adjust easily to misfortune or change}. Hinging upon the general dictionary definition and after reviewing the multiple definitions of resiliency and other close notions in networking, we felt the need to define it more precisely. In our case, with the security of routing functionalities in mind, we define the resiliency as follows:

\begin{defi}[Resiliency] \label{defi1}
Resiliency is the ability of a network to {\it continue to operate} in presence of $k$ compromised nodes, i.e. the capacity of a network to endure and overcome internal attacks. 
\end{defi}

More precisely, it means to achieve a graceful degradation in a packet delivery rate with increasing number of compromised nodes.
In the literature, several conceptually similar properties such as survivability \cite{Ellison1999}, robustness \cite{Sterbenz2002} and resiliency \cite{Wagner2004,Ganesan2001,LiYang2006}, have been discussed. 

The main definition of survivability in information systems is defined in \cite{Ellison1999} as the ability of the network computing system to provide essential services in the presence of attacks and/or failures, and recover full service in a timely manner. Survivability is conceptually similar to resiliency, but from our standpoint, this definition is not precise enough . The main differences between survivability compared to our definition of resiliency are that we insist on the internal attacks when some portion of legitimate nodes is compromised, and we emphasize the network's capacity to endure and overcome these internal attacks.  

In \cite{limaSurvey09}, the authors discuss the survivability by connecting it with intrusion tolerance \cite{deswarte2006}. The authors claim that survivability should be reached by use of preventive, reactive and tolerant approaches operating together. However, this view of survivability corresponds to general security issues, and it is not in the direction that we aim to contribute, i.e. ``beyond cryptography'' approaches. 

Robustness is defined in \cite{Sterbenz2002} as the requirement to accommodate hardware and software failures, asymmetric and unidirectional links, or limited range of wireless communication. It includes the need for the networks to survive specific types of device overrun (physical seizure), network fragmentation and denial-of-service attacks.
The definition of robustness is mainly focused on the failures of the system and does not reflect the network which endures and overcomes the shock caused by internal intruders. 

Other definitions of resiliency were used in several contexts such as data aggregation \cite{Wagner2004}, route failures \cite{Ganesan2001}, key distribution and management \cite{LiYang2006}. According to \cite{Wagner2004} an aggregation function $f$ is $(k; \alpha)$-resilient (with respect to a parameterized distribution $p(X_i | \theta))$ if $rms^*(f; k) \le \alpha \times rms(f)$ for the estimator $f$. The $rms^*(f; k)$ denotes the root-mean-square error of the most powerful $k$-node attack possible. Roughly speaking, an aggregation function $f$ is $(k, \alpha)$-resilient if, for small values of $\alpha$, it can be computed meaningfully and securely in the presence of up to $k$ compromised nodes. This definition is conceptually very close to ours. As \cite{Wagner2004} compares the resiliency of aggregation functions, our aim is to compare the resiliency of routing protocols. In \cite{Ganesan2001} it is argued that the resiliency of a scheme measures the likelihood that, when the shortest path has failed, an alternate path is available between source and sink. This definition focuses on the resiliency of route failures and it does not specifically deal with security issues. Finally, in \cite{LiYang2006} resiliency is defined with respect to cryptographic primitives, which are out of scope of this report. 

\subsection{Network assumptions and Adversary models} \label{models}

In this section we explain the node compromise distribution models and the implemented adversary models. We deal with the security of routing protocols in WSNs and we do not deal with usual cryptographic protections for integrity, confidentiality, authentication and non-repudiation. 

\subsubsection{Network assumptions} \label{network_assumptions}

We consider physically identical sensor nodes and they have the same transmission range. We consider one data collector, called ``sink''. The sensor nodes are densely deployed in a region of size $N \times N$ to collect and transmit data of the physical world to the ``sink''. We define the following traditional assumptions:
\begin{itemize}
\item the ``sink'' is considered robust, having enough resources in terms of memory, computational power and battery to support the cryptographic and routing requirements of WSNs. Thus, adversaries cannot compromise the sink in limited time. 

\item the ``sensor node'' has limited resources in terms of memory, computational power and battery. We assume that sensor nodes are non trustworthy since they are vulnerable to physical attacks and an adversary can compromise them.  
\end{itemize}

A connected graph as the physical topology of the network is considered. The packets are routed from the source to the destination on this physical topology. A fixed radius random graph, which is a common and practical graph model proposed for modeling WSNs, is used. Let us consider a graph $G(\Omega, E )$ where $\Omega$ is a set of nodes wirelessly connected pairwise by a set $E$ of undirected edges representing communication links between nodes.  In this model, the nodes are randomly placed in a $N \times N$ region according to a uniform distribution. A link exists between two nodes $i$ and $j$ if the Euclidean distance between these two nodes is less than the communication range. We assume that the wireless links in our graph are bi-directional, i.e. if node $i$ hears node $j$ then node $j$ also hears node $i$. 

\subsubsection{Adversaries Definitions} \label{adversaries} 

In this section some definitions, described in \cite{rfc4948}, concerning attacks are presented. An attack is an intentional act by which an entity attempts to evade security services and violate the security policy of a system. That is, an actual assault on system security that derives from an intelligent threat.

Attacks can be characterized according the adversaries capacities:
\begin{itemize}
\item  a ``laptop class attacker'' may have access to powerful devices with more computational resources, such as laptops or their equivalent. A single laptop-class attacker might be able to eavesdrop and to jam the entire network.
\item  an ``mote class attacker'' has access to a few motes with the same capabilities as other ordinary sensor nodes. They have no resource advantages over legitimate nodes.
\end{itemize}
			
Attacks can be characterized according to intent:
\begin{itemize}
\item a ``passive attack'' attempts to learn or make use of information from a system but does not affect system resources. For example, a passive eavesdropping that gathers information can compromise the privacy and confidentiality. 
\item an ``active attack'' attempts to alter system resources or affects system operations. Compared to passive attack, the goal of the active attacker is to produce DoS attacks, to disrupt communication by destroying links, to exhaust available resources such as bandwidth or energy etc. 
\end{itemize}
		
Attacks can also be characterized according to point of initiation:
\begin{itemize}
\item an ``outsider attack'' is initiated from outside the security perimeter, by an unauthorized or illegitimate user of the system (an ``outsider''). Numerous external attacks such as jamming, eavesdropping, injecting replayed or fabricated messages can be stated. 
\item an ``insider attack'' is one that is initiated by an entity inside the security perimeter (an ``insider''), i.e. an entity that is authorized to access system resources but uses them in a way not approved by the party that granted the authorization. Selective forwarding, Sybil attacks or Black-Grey-Worm-Sink hole attacks can be mentioned.
\end{itemize}

In our model, the ``mote-class'' attackers are considered, where ordinary sensor nodes can be captured and compromised by an adversary. We deal with an ``insider'' adversary who is ``active''. In Section \ref{implementedAttacks}, the attacks which have been considered for simulations are described in detail. 

\subsection{WSN routing protocols} \label{routingprotocols}

WSNs share some common points with Ad-hoc networks such as lack of infrastructure, decentralized architecture, self-organized and self-configured, radio communication. Thus, Ad-hoc routing techniques greatly inspire WSNs. However, due to specific characteristics (convergecast traffic profile, strong energy constraint, large number of nodes, high node density) some Ad-hoc routing techniques are not suitable for WSNs. 
	
\subsubsection{Classification} \label{classification}

\begin{figure}[ht]
\begin{centering}
        \includegraphics[scale=0.6]{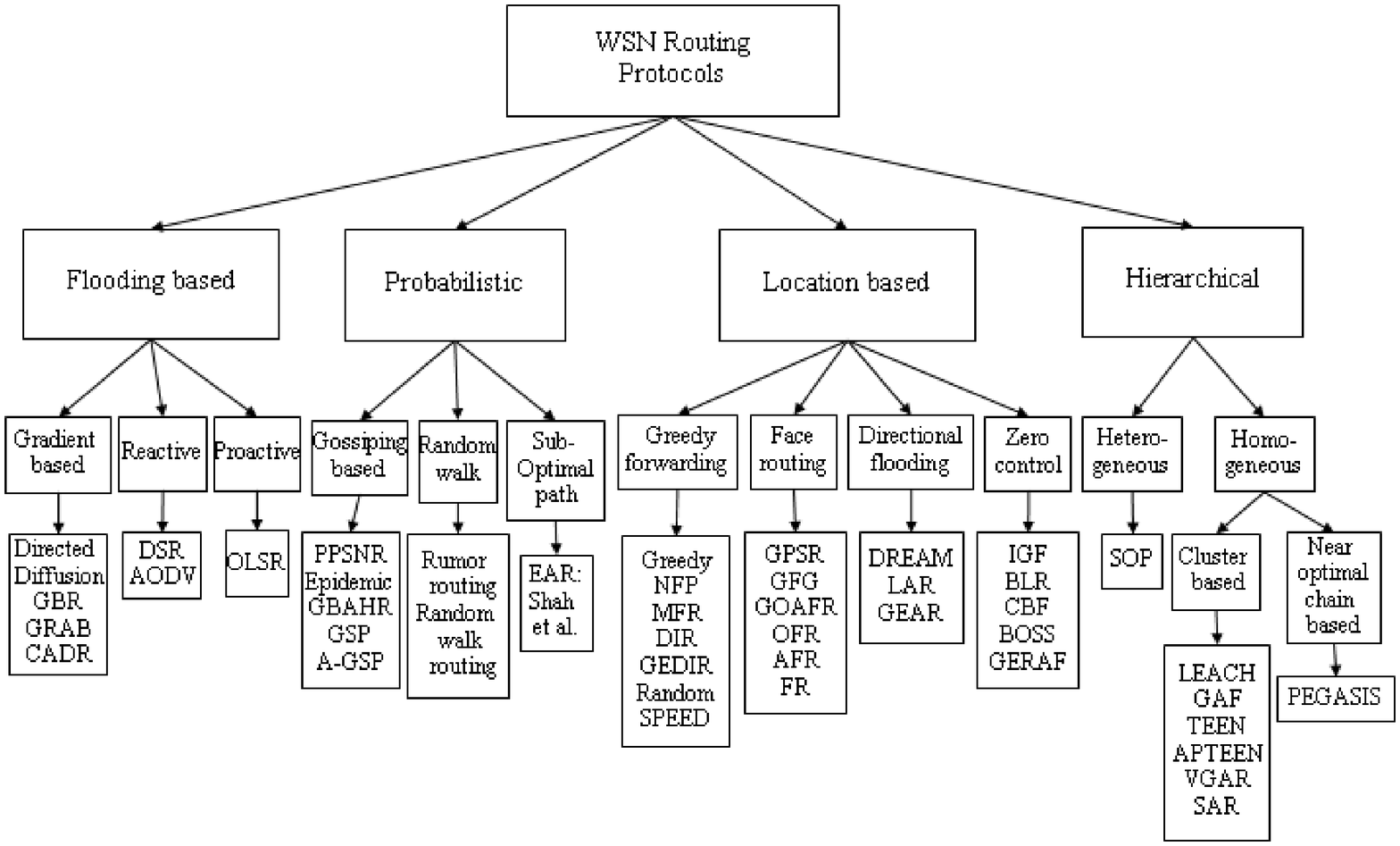} \\
\end{centering}
\caption{Classification of routing protocols in WSNs. } \label{classement}
\end{figure}

We propose a classification of routing protocols in WSNs into four groups; flooding based routing, probabilistic routing, location based routing and hierarchical routing (Fig. \ref{classement}). 

\begin{itemize}
\item Flooding based routing \\

In flooding, a sensor node sends packets to all its neighbors, and neighbor nodes forward them to their neighbors until all nodes are reached. With an ideal MAC layer all nodes can be reached, but in realistic conditions many collisions will occur and there are usually lots of retransmissions, redundancy and packet loss. The common point of the protocols in the flooding category is using a flooding mechanism to discover routes, or to maintain topological information or to setup a gradient metric. 

\item Probabilistic routing \\

Protocols in this category choose the next hop using a dynamically assigned probability or random choice. In the gossiping protocol, a sensor sends packets to a randomly selected neighbor which does the same until destination reached or packet time-to-live expires ($TTL=0$). Gossiping nodes may forward packets back to the sender creating potential inefficiencies and delay problems. The common point of these protocols is the use of some random choice which makes their behavior non deterministic. 

\item Location based routing \\

The common point of the protocols in this category is to use for routing purposes some information about geographical location. Each node has to know the destination node's geographical location, its own location and the location of all its neighbors

\item Hierarchical routing \\

In this category, routing is based on hierarchy of nodes; some nodes can be physically different. Sensor nodes can have different roles such as simple sensors or leader nodes.  Leader nodes can have some special responsibilities such as data aggregation, data fusion, routing whereas simple nodes just sense and transmit data to leader nodes. 

\end{itemize}

\subsubsection{Protocols Under Study} \label{implemented}

To compare the behavior of different routing protocols, under different adversary models, we have selected four routing protocol candidates covering three of the four categories: Dynamic Source Routing (Flooding), Gradient-Based Routing (Flooding), Greedy Forwarding (Geographical) and Random Walk Routing (Probabilistic). 	
	
\begin{itemize}
\item Dynamic Source Routing (DSR)\\

DSR \cite{dsr96} is a flooding based routing protocol, which uses three types of packets: RREQ, RREP and DATA packets. The RERR packet type is not considered in this report. The source node floods a RREQ packet in order to discover routes toward an intended destination (Fig. \ref{dsr1}). Full source-routes are aggregated in the RREQs, and are sent back to the source in RREPs by the sink (Fig. \ref{dsr2}). Once, the source node has received an RREP packet, it updates its routing table and then uses this information to send DATA packets to the sink (Fig. \ref{dsr3}). The route discovery process is provided here only the first time when a source node needs to send a DATA packet to the sink.  

\begin{figure}[ht]
\centerline{
        \subfloat[]{\includegraphics[scale=0.5]{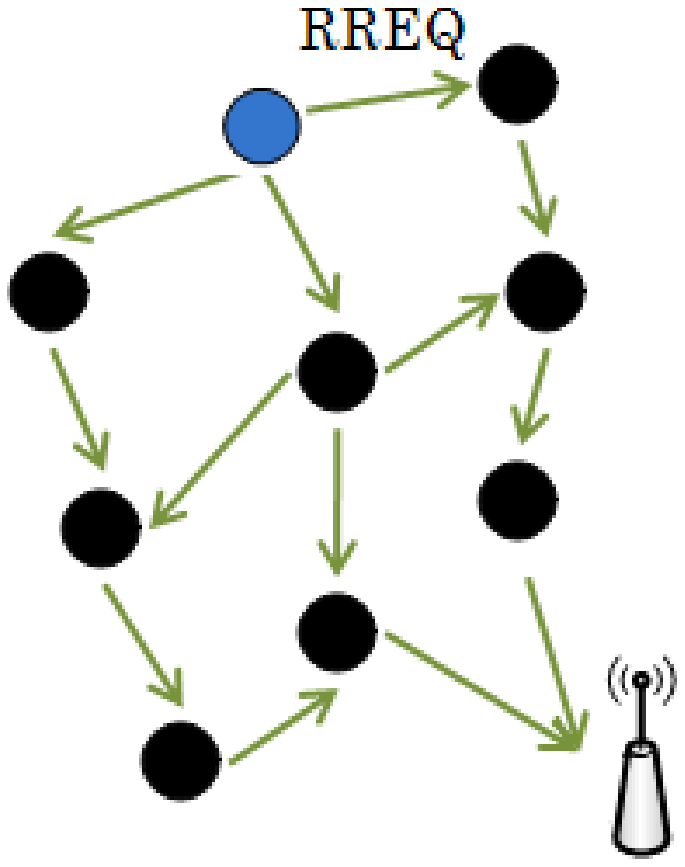} 
         \label{dsr1}}
         \hfil
         \subfloat[]{\includegraphics[scale=0.5]{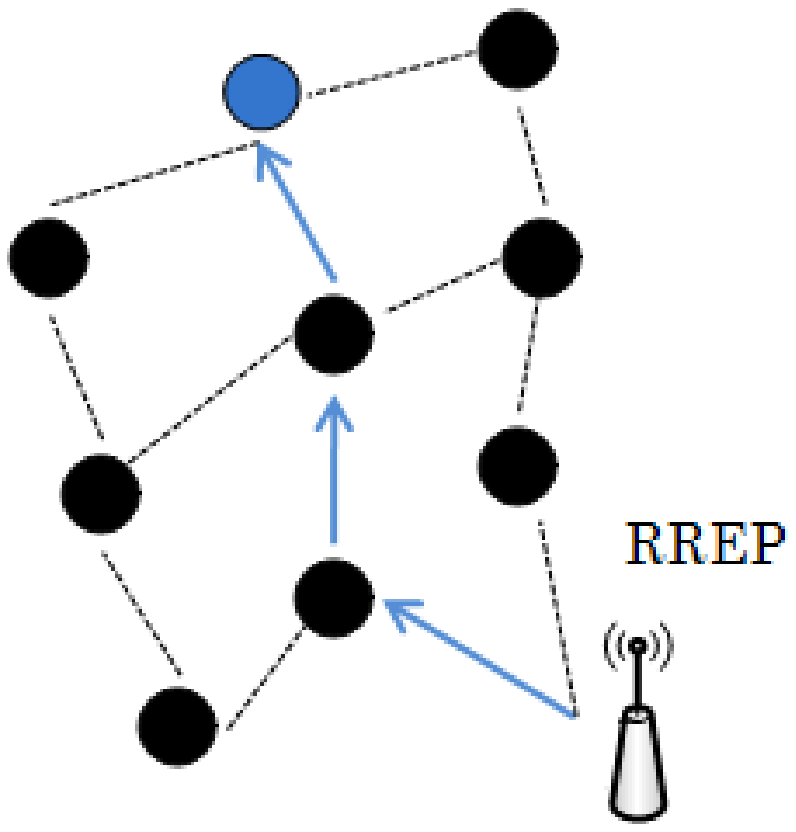} 
         \label{dsr2}}
         \hfil
         \subfloat[]{\includegraphics[scale=0.5]{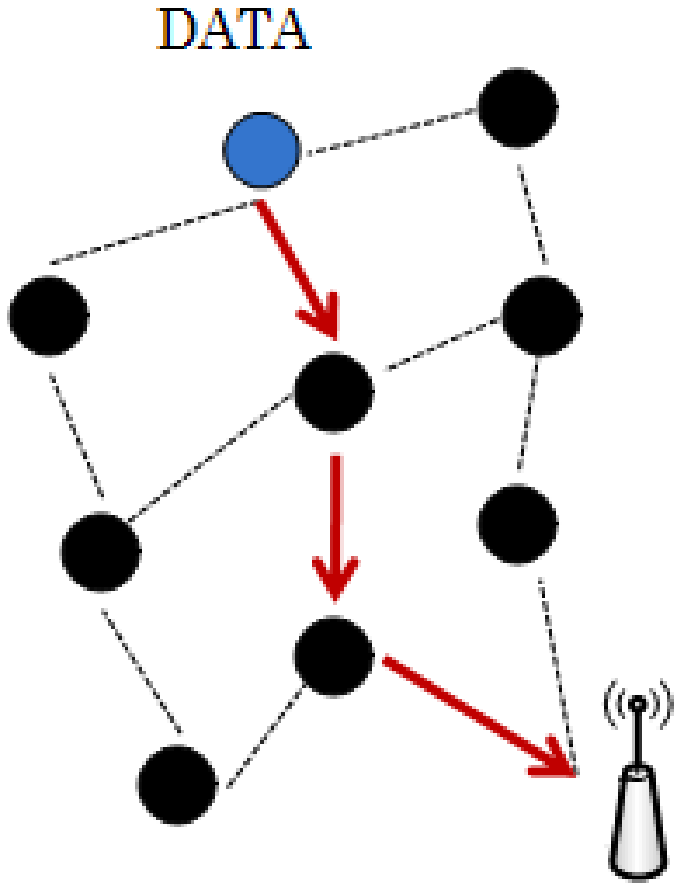} 
         \label{dsr3}}
         \hfil}
\caption{Dynamic Source Routing. (a) Route request phase. (b) Route reply phase. (c) Data dissemination.} \label{dsr}
\end{figure} 

\item Gradient-Based Routing (GBR) \\

GBR \cite{gbr01} is a flooding based routing protocol which uses two types of packets: INTEREST and DATA packets. The sink floods an INTEREST packet in order to setup a gradient (Fig. \ref{gbr1}). The INTEREST packet records the number of hops taken from the sink. Then a node can discover its minimum number of hops from the sink, called the node ``height''. The height difference between a node and one of its neighbors is the gradient on that link. The gradient setup process is provided here only once at the beginning of the simulation. Then nodes send their DATA packets to one of their minimum gradient neighbors and their neighbors do the same until the sink is reached (Fig. \ref{gbr2}).  

\begin{figure}[ht]
\centerline{
        \subfloat[]{\includegraphics[scale=0.5]{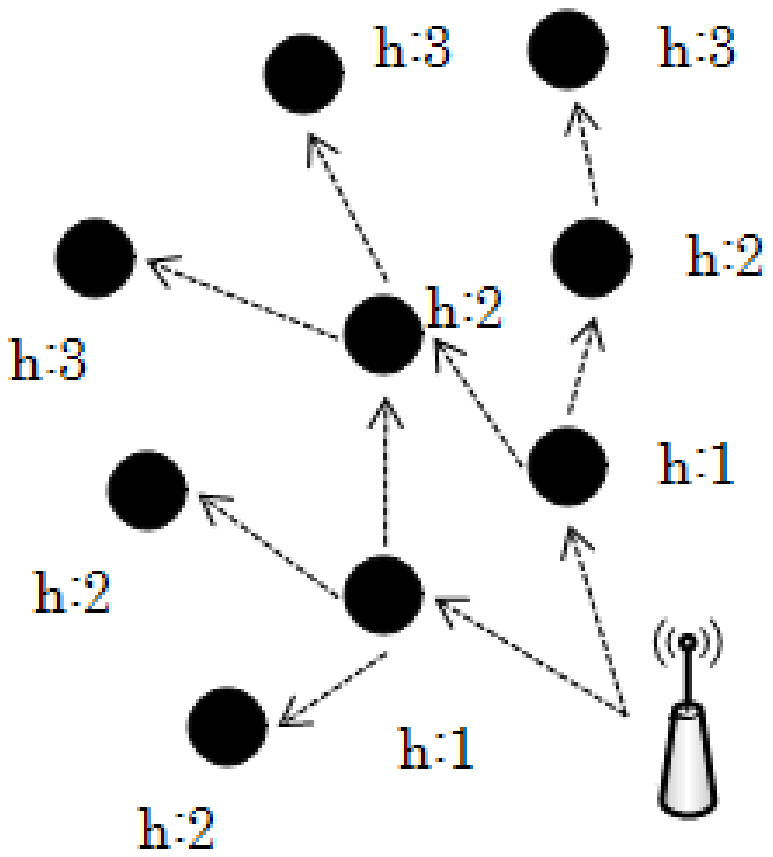} 
         \label{gbr1}}
         \hfil
         \subfloat[]{\includegraphics[scale=0.5]{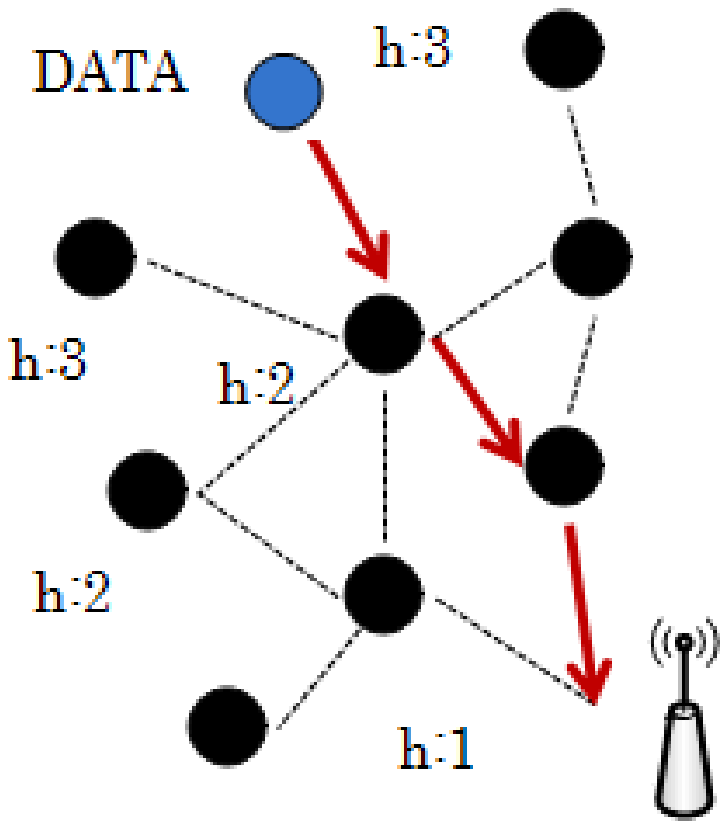} 
         \label{gbr2}}
         \hfil}
\caption{Gradient-Based Routing. (a) Gradient setup phase. (b) Data dissemination.} \label{gbr}
\end{figure}
 
\item Greedy forwarding (GF) \\

Greedy forwarding \cite{gpsr00} is a geographical routing protocol, which uses two types of packets: DATA and HELLO packets. Each node knows its own location and the location of the sink. Each node broadcasts HELLO packets with its identity and location information. All neighbors who receive HELLO packets update their neighborhood table. Each node forwards DATA packets to the neighbor geographically closest to the destination, thus achieving the maximum progress toward the destination (Fig. \ref{greedy}). Here the basic greedy forwarding, without any hole bypassing mechanism, is considered since we use densely and uniformly deployed topologies without holes. 

\begin{figure}[ht]
\begin{centering}
        \includegraphics[scale=0.5]{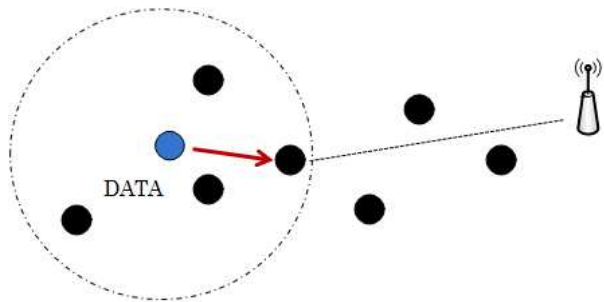} \\
\end{centering}
\caption{Greedy forwarding. Data dissemination. } \label{greedy}
\end{figure}
 
\item Random Walk Routing (RWR) \\

RWR \cite{rwr02} is a probabilistic routing protocol, which uses two types of packets: DATA and HELLO packets. A very simple random walk routing protocol is considered. First, each node broadcasts a HELLO packet with its identity. All neighbors who receive a HELLO packet update their neighborhood tables. Each node sends DATA packets to a randomly selected neighbor, who does the same until the destination is reached or the TTL of the packet expires (Fig. \ref{rwr}). Note that RWR can yield routing loops. 

\begin{figure}[ht]
\begin{centering}
        \includegraphics[scale=0.5]{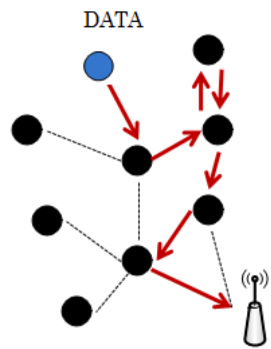} \\
\end{centering}
\caption{Random Walk Routing. Data dissemination. } \label{rwr}
\end{figure}
\end{itemize}
 
\section{Simulations results} \label{simulations}
In this section the simulation setup along with our basic assumptions are stated first in order to subsequently present the obtained simulation results and their analysis.

Simulations were performed using the WSNet \cite{Elyes2008}, which is an event-driven simulator for wireless networks. In the rest of this report, we assume a unique trustworthy sink and untrustworthy sensor nodes since they are vulnerable to physical attacks and can be compromised. 

\subsection{Implemented Attacks} \label{implementedAttacks}

In multi-hop routing, packets may traverse many hops before reaching their destination. The attack which is common to all protocols is DATA packet oriented selective forwarding. According to information used by each protocol, protocol specific attacks such as HELLO and CONTROL packet oriented attacks are also considered.

In the DATA packet selective forwarding attack, malicious nodes simply drop certain messages instead of forwarding all of them. We focus our simulations on two particular scenarios based on selective forwarding for two models for compromised node distribution: a uniform distribution across the whole network area (Scenario 1) and within a smaller area around the sink (Scenario 2). 

In the HELLO packet oriented attack, malicious nodes lie about their identities by claiming illegitimately multiple identities (Sybil). In our case, this attack is considered for RWR and GF, which use HELLO packets in order to establish neighborhood relationship (Scenario 3). GBR and DSR use flooding mechanism to establish their routes and thus are immune to this attack. 

In CONTROL packet oriented attack, malicious nodes introduce false control packets to attract more traffic in order to either exploit them for own needs and/or to drop them with the intention to disrupt efficiently the delivery of data. We considered this attack for DSR and GBR (Scenario 4). 

\begin{figure}[ht]
\begin{centering}
        \includegraphics[scale=0.7]{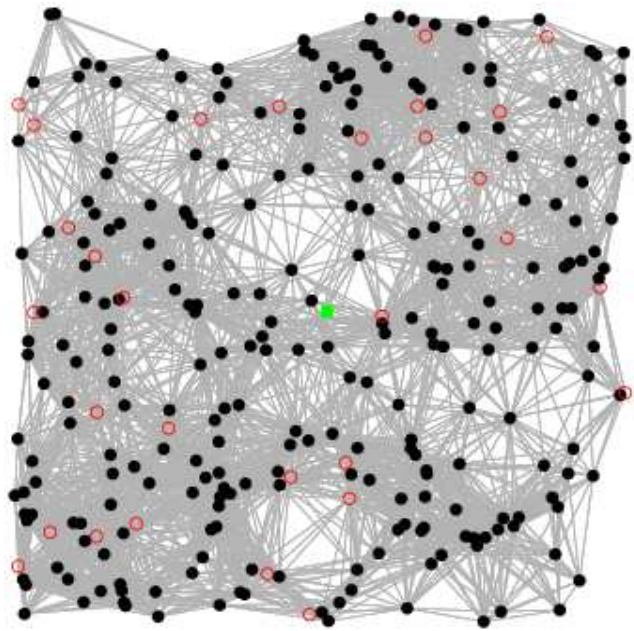} \\
\end{centering}
\caption{Uniformly distributed across the whole network.} \label{uniform}
\end{figure}

\textbf{Scenario 1: DATA packet oriented selective forwarding attack with uniformly distributed compromised nodes across the whole network area.} In this model, we suppose that an adversary has no information about the location of the sink leading to compromised nodes distributed at random positions. Thus, $k$ compromised nodes ($10\%$ to $50\%$ of node population) are randomly and uniformly distributed on a $N \times N$ square field. An example is given in Fig. \ref{uniform} where compromised nodes are shown in red. These malicious nodes drop all DATA packets coming from their neighbors (forwarding probability $p_f=0$). However, they generate and send their own DATA packets to the sink.

\begin{figure}[ht]
\begin{centering}
        \includegraphics[scale=0.7]{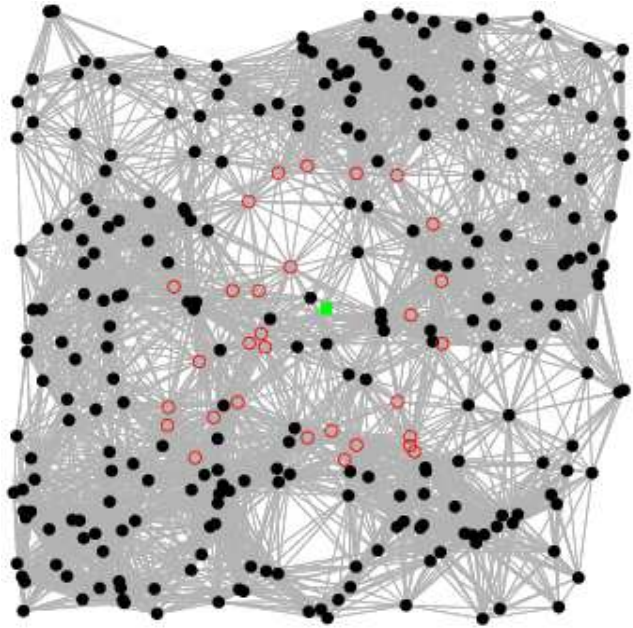} \\
\end{centering}
\caption{Uniformly distributed around the sink.} \label{spatial}
\end{figure}

\textbf{Scenario 2: DATA packet oriented selective forwarding attack with uniformly distributed compromised nodes around the sink (Sinkhole).} In this model, we suppose that an adversary has some information about the location of the sink and he tries to compromise nodes close to the sink. Thus, a square of size $M \times M$ is defined around the sink, which is $1/4$ of $N \times N$, and $k$ compromised nodes are randomly distributed in it as shown in Fig. \ref{spatial}. The compromised nodes (shown in red) behave as described in scenario 1. 

\textbf{Scenario 3: HELLO packet oriented attack with uniformly distributed compromised nodes across the whole network area.} In this model, $k$ compromised nodes are randomly and uniformly distributed on a $N \times N$ square field as given in Fig. \ref{uniform}. In our case, all protocols send periodically HELLO packets to establish neighborhood relationships. Compromised nodes lie about their identities by claiming illegitimately multiple identities (Sybil). 

In this adversary model, malicious nodes introduce false HELLO packets by producing a new identity for each sent. A HELLO packet is sent every $3$ seconds. Every node purges its neighborhood table every $7.5$ seconds. For each periodic sent of HELLO packet, the compromised nodes choose a new identity in the interval $[0, N]$, where $N$ is the total number of nodes. The new identity is different from its real identity and the identity of its direct neighbors. We refer to a malicious device's additional identities as Sybil nodes. According to Sybil attack taxonomy \cite{Newsome2004}, our model corresponds to ``direct communication'' where Sybil nodes communicate directly with legitimate nodes, ``fabricated identities'' where an attacker can simply create arbitrary new Sybil identities and ``non-simultaneous'' form where an attacker might present a large number of identities over a period of time, while only acting with a smaller number of identities at any given time.

\textbf{Scenario 4: CONTROL packet oriented attacks with uniformly distributed compromised nodes across the whole network area.} In this model, $k$ compromised nodes are randomly and uniformly distributed on a $N \times N$ square field as given in Fig. \ref{uniform}. Compromised nodes introduce false CONTROL packets to attract more traffic. 

These attacks are applicable to protocols using CONTROL packets (RREQ, RREP, INTEREST etc.). In our case, we considered false RREP packets for DSR and false INTEREST packets for GBR. We do not consider false RREQ packets in DSR, because for malicious nodes there is no interest. In GBR, when the sink node floods INTEREST packet in order to setup gradients, compromised nodes modify the number of hops which results in claiming to be nodes with better gradient values. In DSR, when a source node floods a RREQ packet, to discover a route to the sink node, compromised nodes intercept the RREQ packet and send a false RREP packet with a false path. In both cases, a malicious node claims to be a node which is a direct neighbor of the sink.

\subsection{Simulation Assumptions and Environment} \label{assumption}

$300$ sensor nodes are randomly and uniformly distributed over a square field of $100m \times 100m$. A unique sink is assumed at the center of the field. The deployed nodes have fixed positions during each simulation. The radio range is $20m$ resulting in an average degree per node of $31$. The time to live (TTL) of each packet is fixed to $32$ hops. 

WSNet gives the possibility to define radio propagation and interfaces to a great level of detail. However, for our purposes on focusing on routing aspects and consider only the impact of the defined attacks on routing, we configure WSNet for an ideal physical/mac layers (e.g. no interference, no path-loss, omni-directional antennas and no collisions). 

Traffic is generated using a Poisson model. Hence, the packet inter-arrival time follows an exponential distribution with parameter $\lambda$, about $1 packet/sec$ per node. The simulation time is $100$ seconds, and the total number of generated packets is about $30 000$. The simulations are averaged over $100$ trials for each adversary model and for each protocol with a $95\%$ confidence interval. Table \ref{sum} summarizes the simulation parameters.
\begin{table}
\begin{center}
\begin{tabular}{|l|c|r|}
  \hline
  Parameter & Value \\
  \hline
  Number of nodes & $300$  \\
  Area size  & $100 \times 100m$  \\
  Transmission range & $20m$ \\
	Topology & uniformly distributed \\ 
	Traffic generation & Poisson distribution $\lambda$ $= 1 p/s$\\
	Number of runs &  $100$ \\
	Simulation time & $100s$ \\
  \hline
\end{tabular}
\caption{Summary of the simulation parameters.} \label{sum}
\end{center}
\end{table}

\subsection{Evaluation Metrics}\label{metrics}	

The main responsibility of the routing layer is to ensure reliability of the network. Reliable delivery of data characterizes the success of routing protocols. To gain insight concerning the WSN routing resiliency, as defined in Definition \ref{defi1}, the following three evaluation metrics are used: 
\begin{itemize}
\item \textbf{Average delivery ratio:} \textit{Delivery ratio = total number of received packets by the sink / total number of sent packets by the sensors.} This is the most important metric in order to evaluate the success of routing functionality and the reliability of the network. Without any attacks and without any interferences and collisions, all DATA packets are received successfully by the sink and the delivery ratio is $1$. The delivery ratio for the four chosen routing protocols is measured for the two scenarios by varying the number of compromised nodes.  
\item \textbf{Average degree of nodes:} \textit{Degree of nodes = number of neighbors for each node.} The average degree for each node is measured in order to detect neighborhood abnormalities. 
\item \textbf{Average path length:} \textit{Path length = number of hops crossed for each received packet.} This metric allows us to determine the number of forwarding nodes on a route. 
\end{itemize}

\subsection{Results and Analyses}\label{results}

\textbf{Scenario 1, Results and Analyses}

\begin{figure*}[t]
\vspace{-2mm}
\centerline{
			\subfloat[]{\includegraphics[scale=0.45]{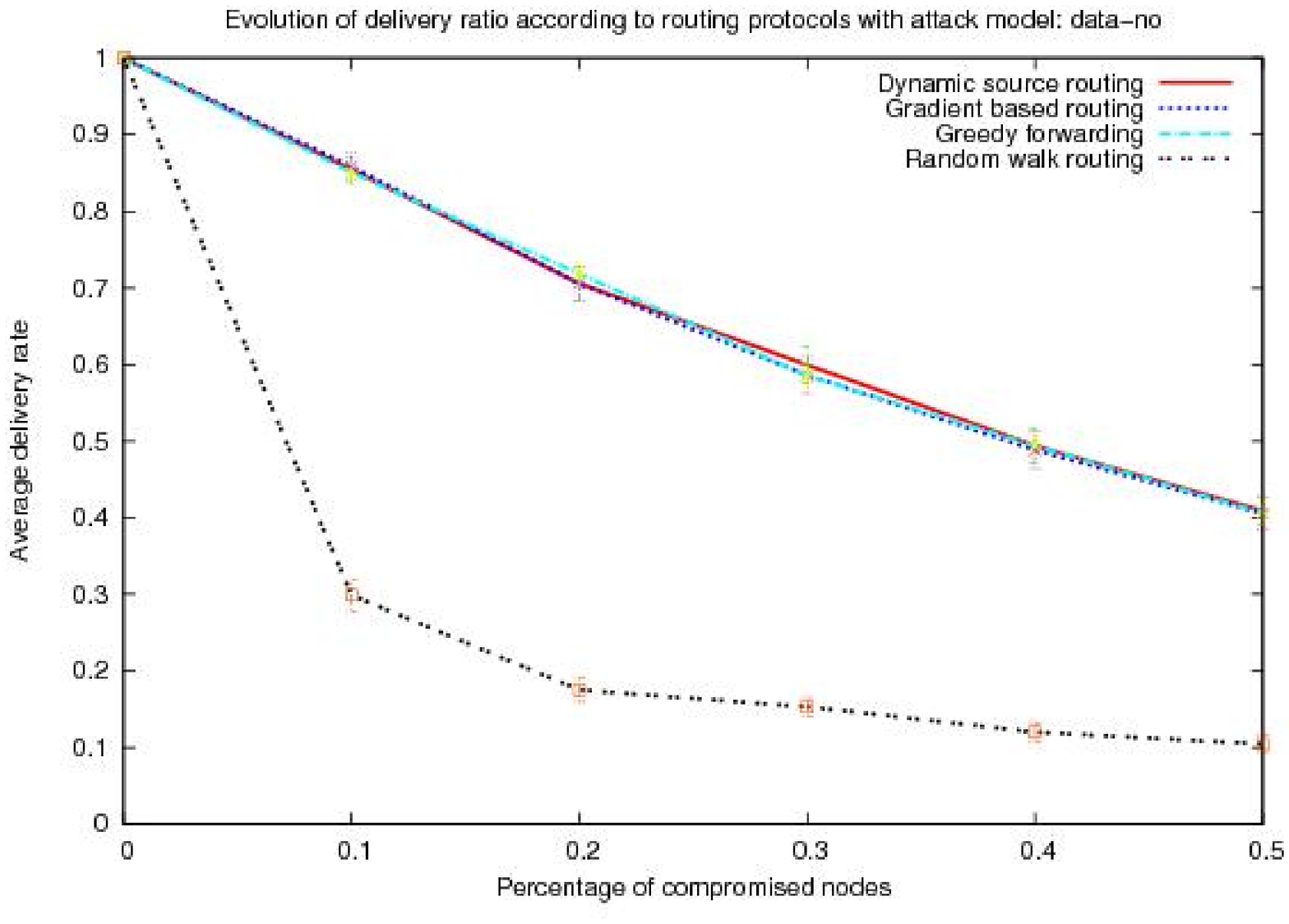}
         \label{delivery1}}
         \hfil
      \subfloat[]{\includegraphics[scale=0.45]{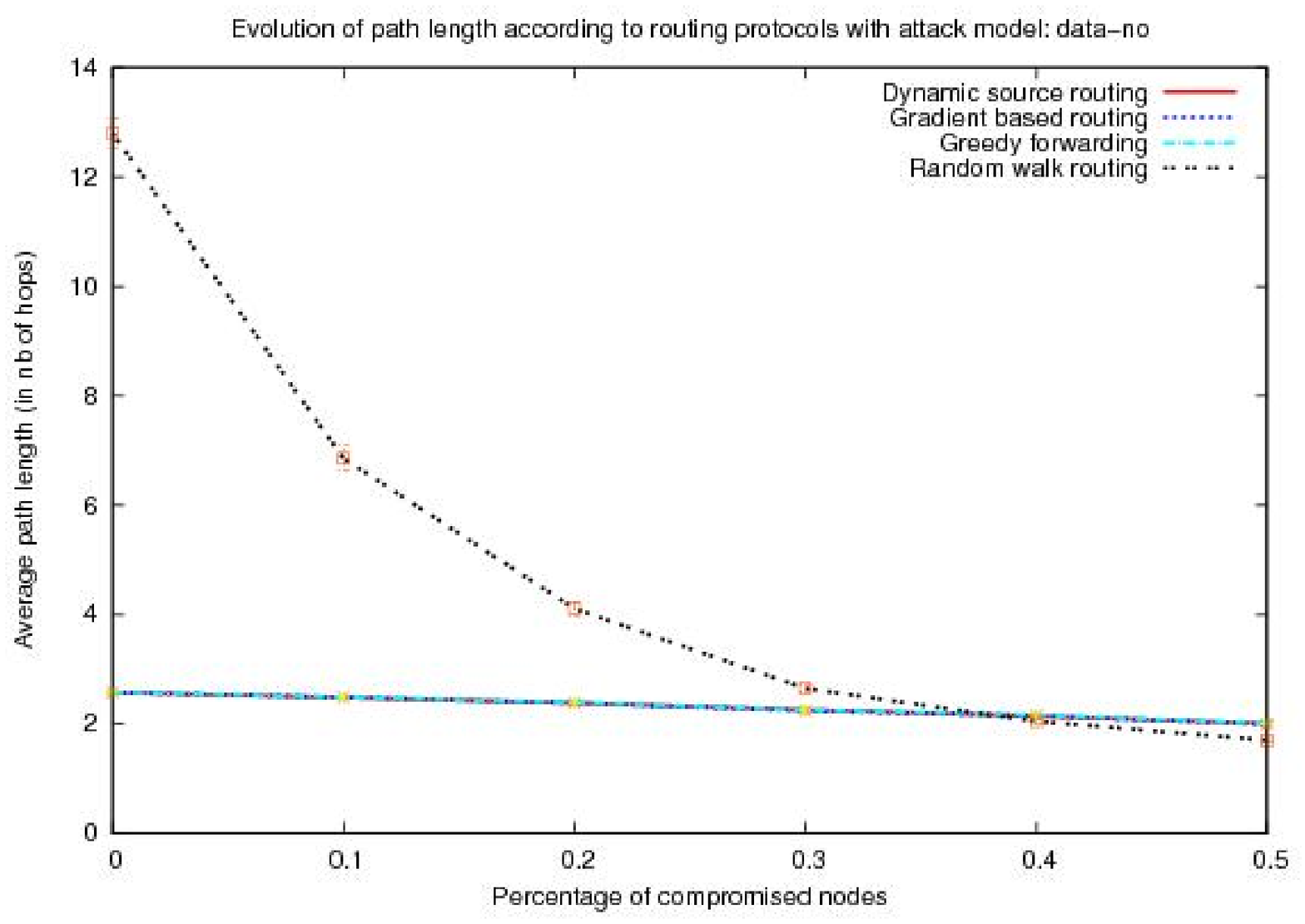}
         \label{path1}}
      	 \hfil 
       }
\caption{Scenario 1. (a) Average delivery ratio. (b) Average path length.}
\label{scenario1}
\vspace{-2mm}
\end{figure*}
 
Fig. \ref{delivery1} shows the average delivery ratio of the four chosen routing protocols (DSR, GBR, GF, RWR) for Scenario 1. As expected the average delivery ratio decreases, when the percentage of compromised nodes increases. GBR, DSR and GF have similar results, whereas, RWR is worst in successful packet delivery. The difference between RWR and the other protocols is due to the method of route choice: RWR will not privilege automatically the shorter paths since each node sends DATA packets to a randomly selected neighbor. Hence, DATA packets can take long routes. 

Fig. \ref{path1} shows the average path length of the four routing protocols for Scenario 1. RWR has much longer path length than others. The path length is inversely proportional to the average delivery ratio. When the path length is high, the number of forwarding nodes is high. Thus, the probability to meet malicious forwarding nodes increases. Moreover, we observe that the path length of RWR decreases, when the percentage of compromised nodes increases. Hence, received DATA packets come mainly from the sensor nodes which are the closest to the sink. This fact can be explained as follows: if $l$ denotes the path length in number of hops from source to destination; $p_c$ denotes the probability that a node is compromised and $p_n$ is the probability that a packet is delivered (i.e. all forwarding nodes on the route are legitimate), we have $p_n = (1 - p_c)^{l}$. The probability to find a ``safe'' route in RWR exponentially decreases with route length.

DSR, GBR, GF choose the next hop closest to the destination, thus DATA packets meet most of the time the same nodes which are at the center of the field. Here we observe that the shortest path strategy leads to a greater delivery ratio: because packets have less probability to meet malicious nodes and only a few number of nodes at the center of the field are exploited.  However, deterministic route choice could be considered as a bad property for resiliency since all packets will be lost if at least one forwarding node on the route is compromised and due to the deterministic choice the structural redundancy of physical topology will not be effectively exploited. RWR in theory is capable of exploiting the potential connectivity of physical topology but suffers from the route length effect.

\textbf{Scenario 2, Results and Analyses}

\begin{figure*}[t]
\vspace{-2mm}
\centerline{
			\subfloat[]{\includegraphics[scale=0.45]{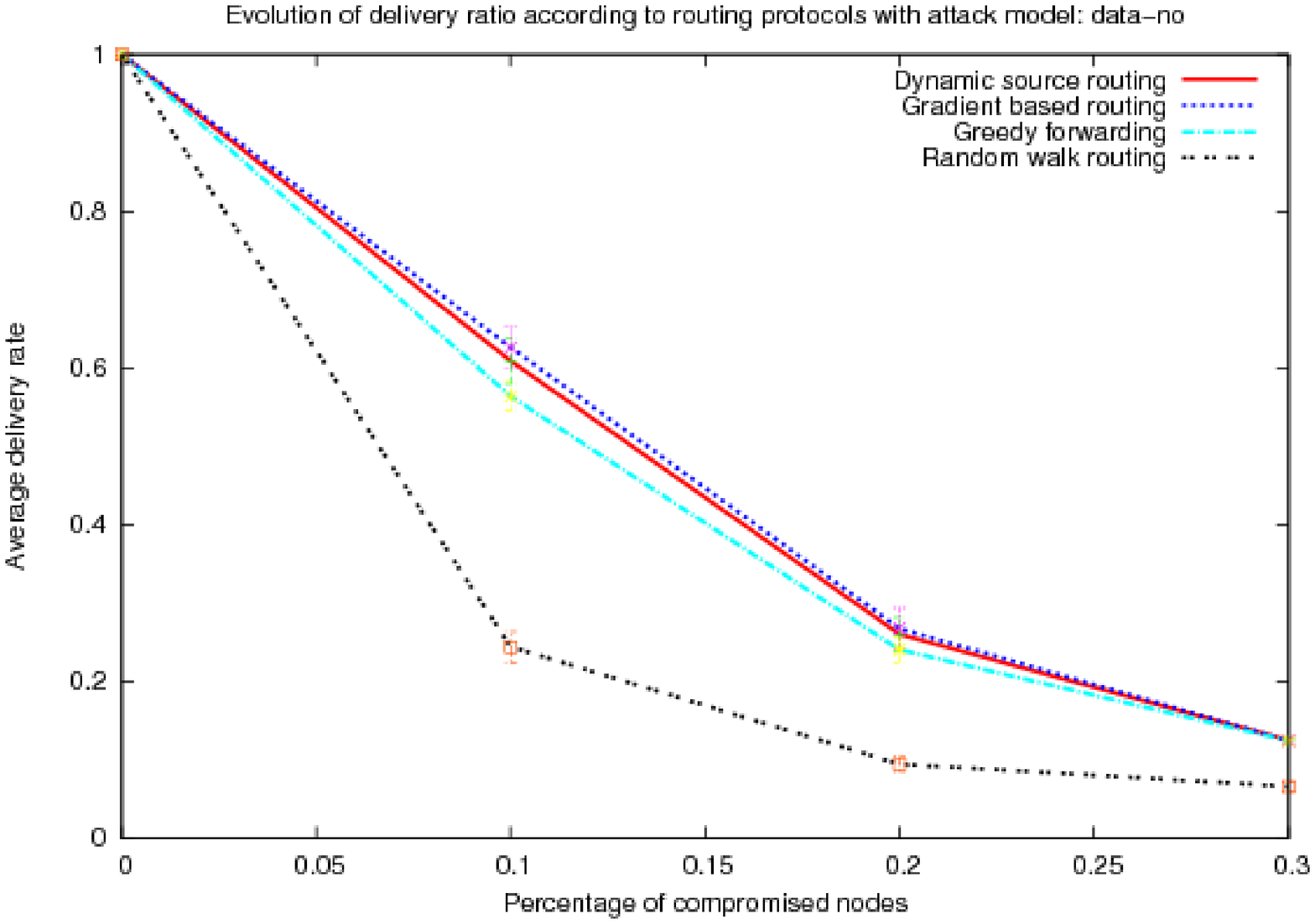}
         \label{delivery2}}
         \hfil
      \subfloat[]{\includegraphics[scale=0.45]{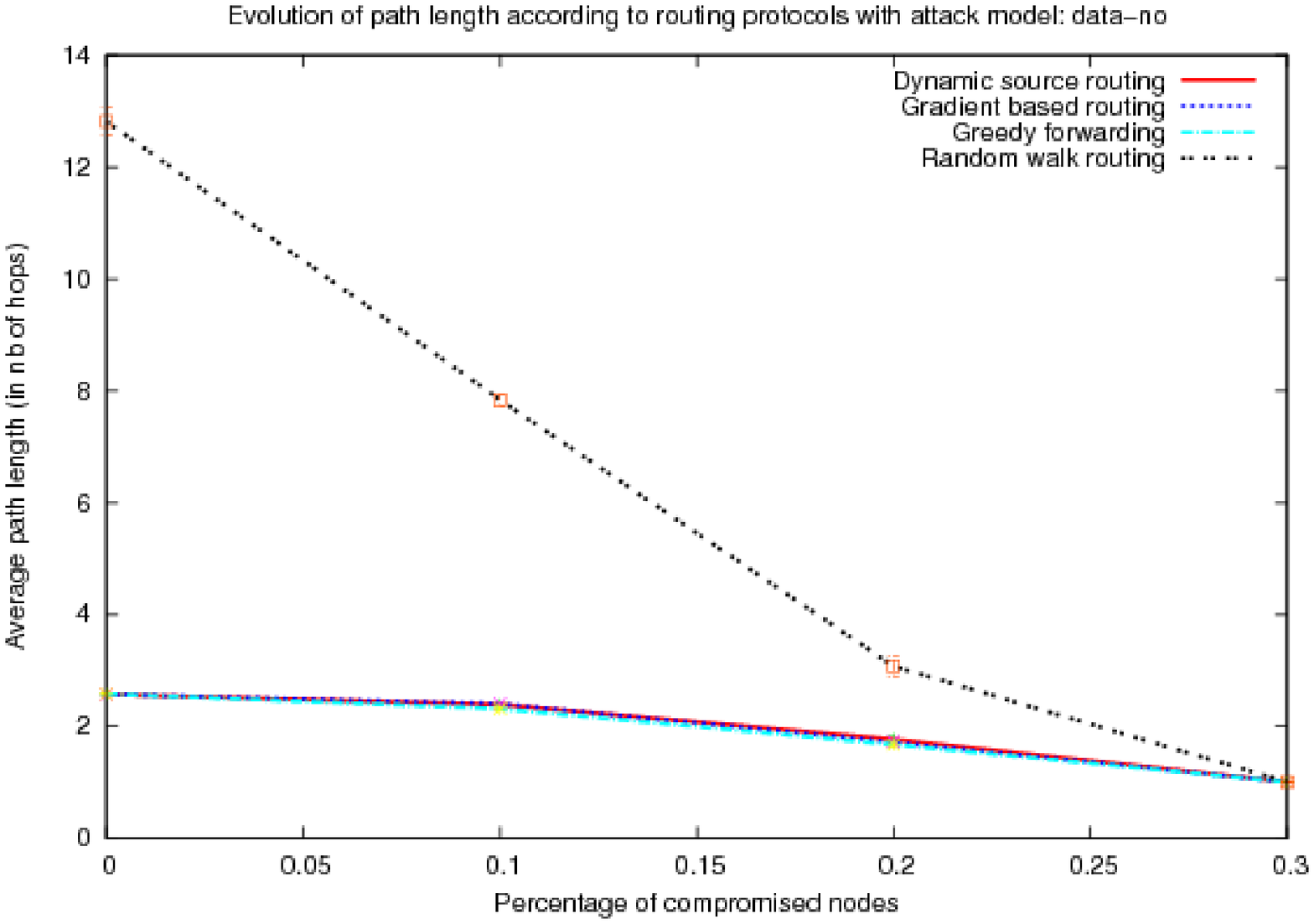}
         \label{path2}}
      	 \hfil 
       }
\caption{Scenario 2. (a) Average delivery ratio. (b) Average path length. }
\label{scenario2}
\vspace{-2mm}
\end{figure*}

Fig. \ref{delivery2} shows the average delivery ratio of the four chosen routing protocols (DSR, GBR, GF, RWR) for Scenario 2. Compared to Scenario 1, the distribution of compromised nodes is localized around the sink leading to a Sinkhole attack. We observe that the impact of attacks is more important than for Scenario 1. When the compromised nodes are close to the sink, they receive for retransmission more packets than other nodes and thus attract most of the traffic creating a ``donut effect'' around the sink. Fig. \ref{path2} shows the average path length of the four protocols for Scenario 2. When all nodes around the sink are compromised, the sink receives packets only from these malicious nodes. No DATA packets are received by the sink from the legitimate nodes. That is why we observe on the Fig. \ref{path2} a path length that tends to $1$.

\textbf{Scenario 3, Results and Analyses}

\begin{figure}[ht]
\begin{centering}
        \includegraphics[scale=0.5]{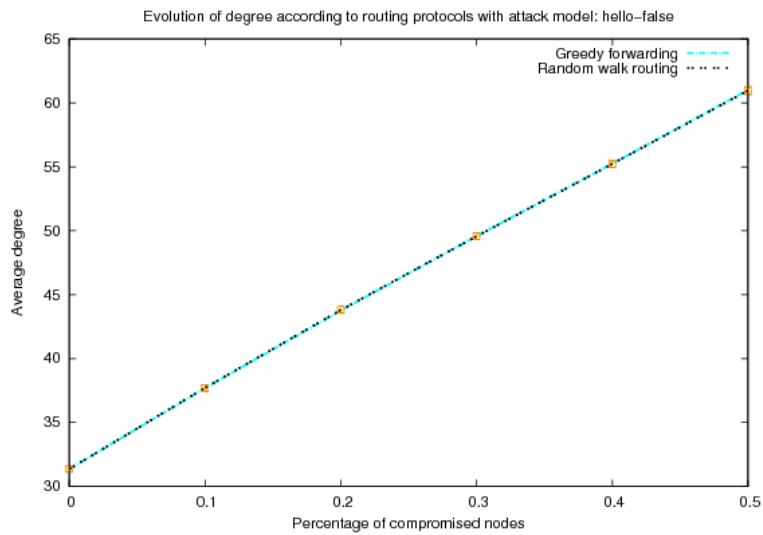} \\
\end{centering}
\caption{Scenario 3. Average degree of nodes.} \label{degreeFalseHello}
\end{figure}

\begin{figure*}[t]
\vspace{-2mm}
\centerline{
	    \subfloat[]{\includegraphics[scale=0.45]{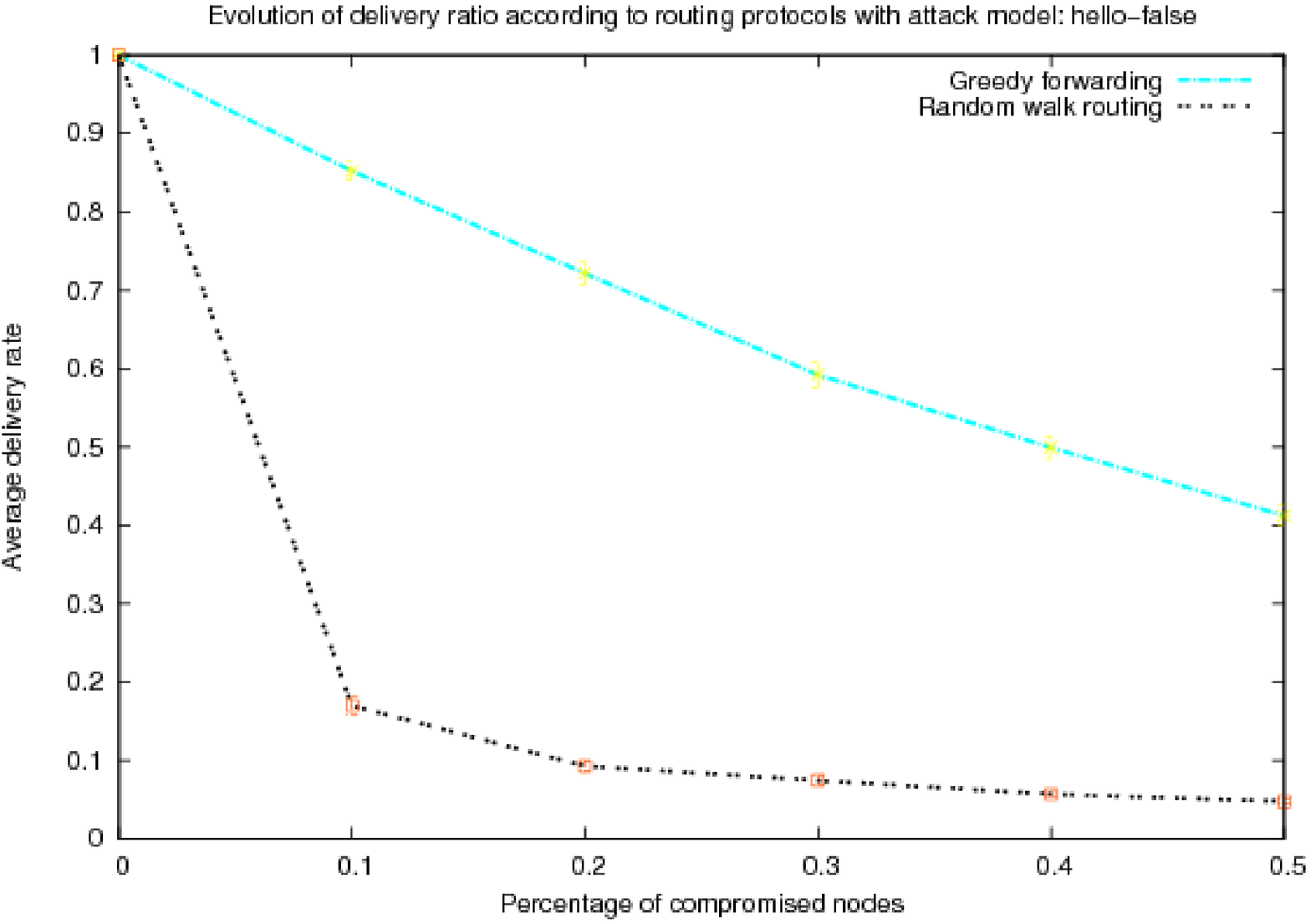}
         \label{deliveryFalseHello}}
         \hfil
      \subfloat[]{\includegraphics[scale=0.45]{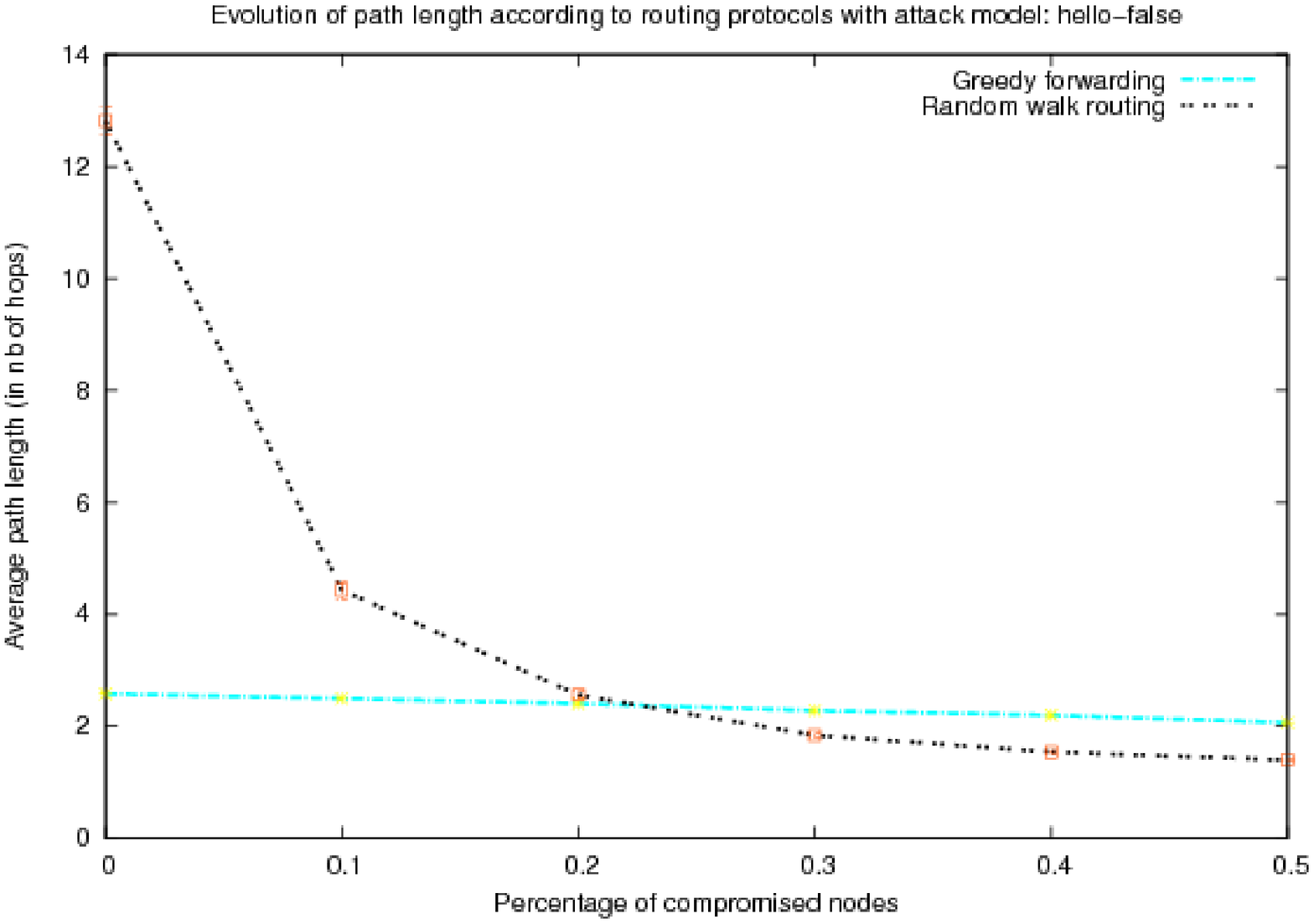}
         \label{pathFalseHello}}
      	 \hfil 
       }
\caption{Scenario 3. (a) Average delivery ratio. (b) Average path length.}
\label{scenario3}
\vspace{-2mm}
\end{figure*}

Fig. \ref{degreeFalseHello} shows the average degree of nodes of GF and RWR for Scenario 3. As expected the average degree increases, when the number of compromised nodes increases. We consider $N = 300$ the total number of network nodes. Lets consider $10\%$ of malicious nodes $k = 30$. \\

$N - k = 300 - 30 = 270$ is the number of legitimate identities. Each malicious node has the time to send $2.5$ false HELLO packets before its neighbors purge their neighborhood tables. The total number of created false identities is $30 \times 2.5 = 75$. Hence, the total number of identities at the network is $270 + 30 \times 2.5 = 345$ instead of $300$. As the average degree of nodes without attack is $32$, when $10\%$ of nodes are compromised, the average degree of nodes is $345 \times (32 \div 300)$, which is near to $37$ as observed on the Fig. \ref{degreeFalseHello}. 

Fig. \ref{deliveryFalseHello} shows the average delivery ratio of GF and RWR for Scenario 3. The average delivery ratio decreases, when the number of compromised nodes increases. For the next hop, GF chooses a neighbor closest to the sink and RWR chooses a neighbor randomly. As the neighborhood table is established with HELLO packets, when the chosen next hop is a not existing false identity (Sybil node), a DATA packet is lost. GF is best in successful packet delivery than RWR due to path length as shown in \ref{pathFalseHello} (the same reason than with Scenario 1). RWR has much longer path than others.

\textbf{Scenario 4, Results and Analyses}

\begin{figure}[ht]
\begin{centering}
        \includegraphics[scale=0.5]{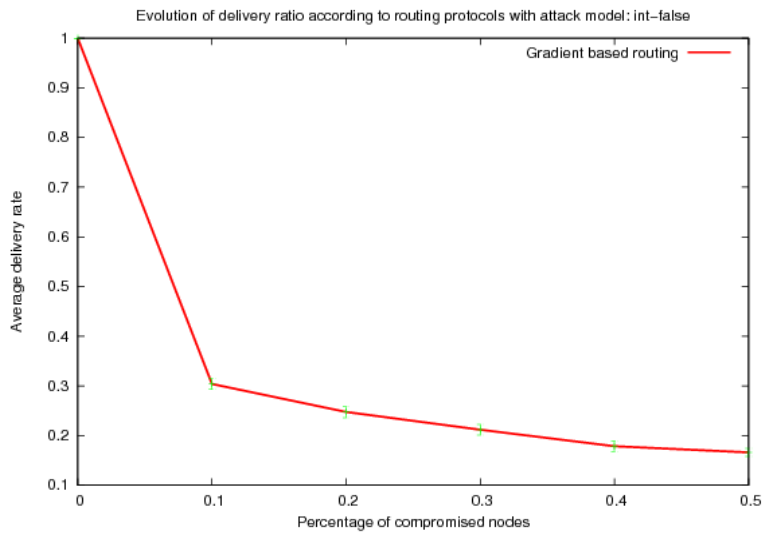} \\
\end{centering}
\caption{Scenario 4. Average delivery ratio of GBR.} \label{deliveryFalseInterest}
\end{figure} 

\begin{figure}[ht]
\begin{centering}
        \includegraphics[scale=0.5]{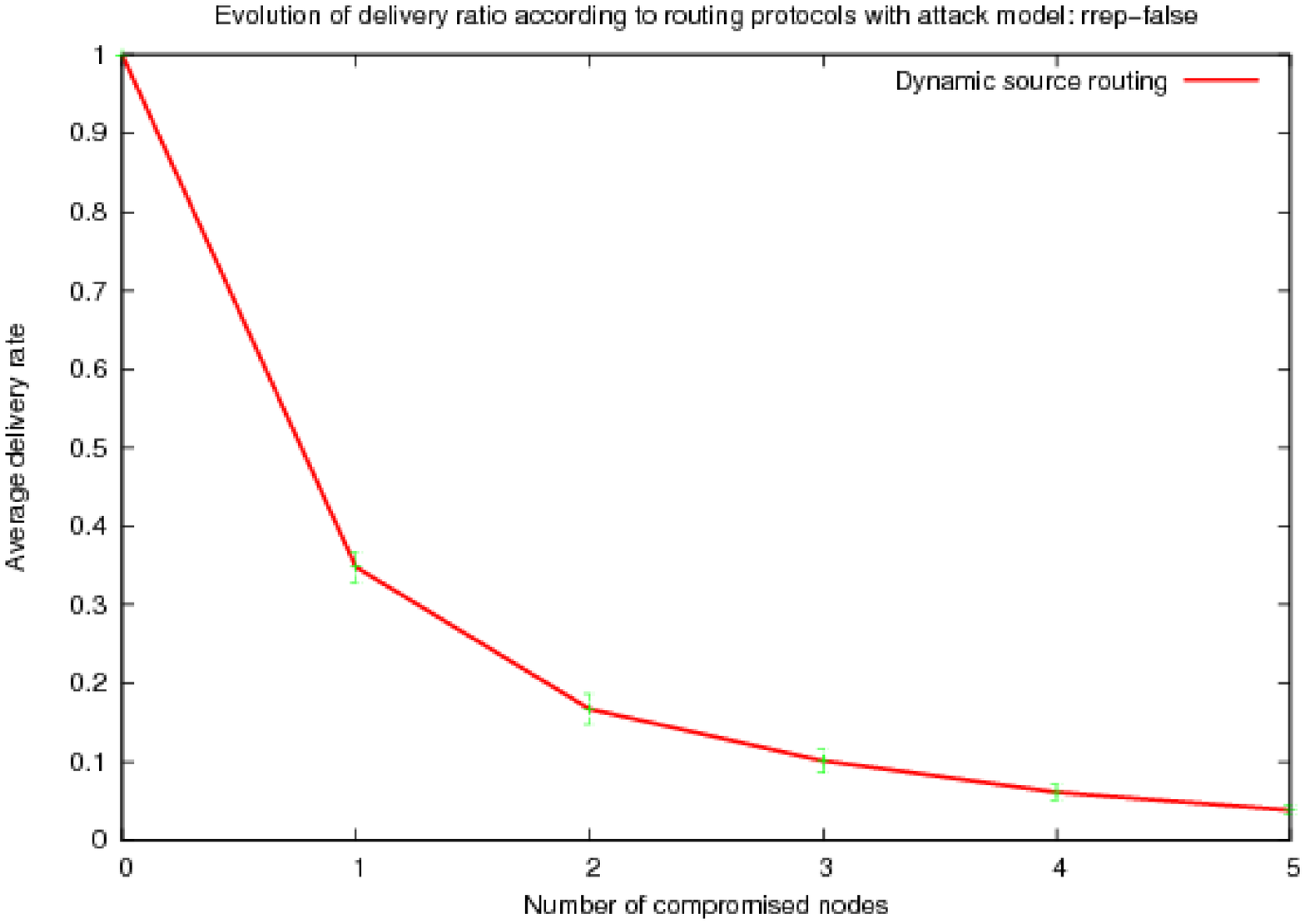} \\
\end{centering}
\caption{Scenario 4. Average delivery ratio of DSR.} \label{deliveryFalseRrep}
\end{figure}

Fig. \ref{deliveryFalseInterest} shows the average delivery ratio of GBR for Scenario 4 with false INTEREST packets. The average delivery ratio decreases, when the percentage of compromised nodes increases. We observe a deeper impact than Scenario 1 as shown on the Fig. \ref{delivery1}. Lets consider $10\%$ of malicious nodes which means $k = 30$. Over two-thirds of the traffic is lost with only $10\%$ of compromised nodes. 
In GBR, the sink floods INTEREST packet in order to setup gradient. In GBR, for the next hop, nodes choose a neighbor which has the better gradient to forward DATA packets. Malicious nodes improperly modify hop count of the INTEREST packet by claiming to be a direct neighbor of the sink. Most of the time, a legitimate node sends DATA packets to a malicious node. Once the traffic is attracted by a malicious node, it drops all DATA packets thereby producing efficient disruption of DATA delivery. 

Fig. \ref{deliveryFalseRrep} shows the average delivery ratio of DSR for Scenario 4 with false RREP packets. The average delivery ratio decreases very quickly, when the number of compromised nodes increases. We observe that the impact of attacks is much more important than a simple selective forwarding attacks (Scenario 1 and 2). 

In DSR, a source node discovers a route toward the sink by flooding RREQ packet. When the sink receives RREQ packet it sends back a RREP packet to the source node. As RREQ packet is flooded, malicious nodes receive also RREQ packets. Then they create a false RREP packet with false path and send it to the source node. In most of the time a legitimate node sends DATA packets to a malicious node. Once the traffic is attracted by a malicious node, it drops all DATA packets thereby producing efficient disruption of DATA delivery.

Here, we consider only a few number of compromised nodes (1 to 5) instead of some percentage ($10\%$ to $50\%$) as considered earlier. Because with $10\%$ of compromised nodes, the sink receives only its direct neighbors' DATA packets. Lets consider one malicious node, which means $k = 1$. The total number of nodes is $300$, including a sink and $299$ source nodes. A single compromised node can impact $197$ legitimate nodes over $299$ and attracts all their DATA traffic. Two-thirds of the DATA packets are lost with only one compromised node. When $k = 2$, the first compromised node impacts $109$ legitimate nodes and the second one impacts $167$. A total of $276$ legitimate nodes over $299$ are impacted with only two compromised nodes. Thus, we observe a significant decrease of average delivery ratio when $k = 2$.  

\subsection{Requirements for Resiliency}

Firstly, we can say that more the protocol is statefull and routing depends on state information the more it will be vulnerable to attacks targeting this information. For instance, flooding based routing protocols such as DSR, GBR are vulnerable to the attacks targeting route discovery process with false CONTROL packets, whereas GF and RWR are vulnerable attacks targeting HELLO packets. The impact of CONTROL packet oriented attack is much more important than others. Only one compromised node can attract two-thirds of the DATA packets. In our point of view, it is important to:

\begin{enumerate}
\item keep WSN routing protocols as stateless as possible to avoid the proliferation of specific attacks and
\item provide for a degree of random behavior to prevent the adversary from determining which are the best nodes to compromise. 
\end{enumerate}

Next, we can distinguish three requirements for resiliency. First, the graph representing WSNs should be connected to get the reliability between source and destination. Second, the degree of the nodes must be high, which increases the number of candidates for next hop and provides for enough structural redundancy. Third, the route must be diversified in order to exploit the structural redundancy of the physical topology while balancing the increase of route length. For example, in flooding based routing such as DSR and GBR, the route diversity depends on their route maintenance phase. It depends on how often the routes are updated and new routes are established. To have better resiliency, we can improve existing deterministic routing protocols such as DSR, GBR, GF by introducing some randomness on their behavior or we can design a protocol which choose different routes randomly for each packet to send. In any case design for resiliency will have an energy cost which is an aspect that needs to be quantified.

Finally, with scenario 2, we observe that an attack can have greater impact if an adversary captures nodes closer to the sink. When an adversary captures all nodes around the sink, it effectively isolates the sink from the rest of the network. It is equivalent to say that the sink is compromised, even if it is considered physically tamper proof. Hence, it becomes important to keep secret the position of the sink. Another solution is to provide for redundant sinks. 

\section{Conclusion} \label{conclusion}

In this report, a preliminary study for WSNs security of the routing layer is presented from the standpoint of resiliency to attacks. First, we presented an overview of security issues for WSNs generally and at the network layer in particular, including existing attacks and defensive measures. Second, a definition of resiliency for routing protocols is defined and compared with other similar notions. Third, a classification of WSN routing protocols is suggested. Finally, we presented simulations and analyses of four particular routing protocols from different categories to determine their resiliency against DATA-HELLO-CONTROL packets oriented attacks accusing three metrics (average delivery ratio, node degree and path length in number of hops). From those analyses, we deduced some requirements at the routing layer to enhance the network resiliency in the face of those attacks.

In the future we intend to give further precision and formality to our definition of resiliency which will permit more precise experimentation and in-depth analysis and quantify the energy costs of resiliency. We plan to define more eloquent metrics incorporating the energy aspect and further explore the interplay of structural (topology) and behavioral (protocol) redundancy in the emergence of resiliency properties.
 
\bibliographystyle{plain}
\nocite{*}
\bibliography{Biblio_RR} 

\begin{thebibliography}{10}

\bibitem{resilience}
{http://www.merriam-webster.com/dictionary/resilience}.

\bibitem{Chan2007}
G.~Noubir A.~Chan, X.~Liu and B.~Thapa.
\newblock Broadcast control channel jamming: Resilience and identification of
  traitors.
\newblock In {\em IEEE International Symposium on Information Theory (ISIT)},
  pages 2496--2500, Nice, France, June 2007.

\bibitem{Durresi2007}
M.~Durresi A.~Durresi, V.~Paruchuri and L.~Barolli.
\newblock Anonymous routing for mobile wireless ad hoc networks.
\newblock {\em International Journal of Distributed Sensor Networks (IJDSN)},
  3(1):105--117, 2007.

\bibitem{Akyildiz2002}
I.~F. Akyildiz, W.~Su, Y.~Sankarasubramaniam, and E.Cayirci.
\newblock Wireless sensor networks: a survey.
\newblock {\em Computer Networks}, 38(4):393--422, January 2002.

\bibitem{Alarifi2006}
Abdulrahman Alarifi and Wenliang Du.
\newblock Diversify sensor nodes to improve resilience against node compromise.
\newblock In {\em Proceedings of the 4th ACM Workshop on Security of ad hoc and
  Sensor Networks (SASN)}, pages 101--112, Alexandria, USA, October 2006.

\bibitem{Alzaid2008}
H.~Alzaid, E.~Foo, and J.~G. Nieto.
\newblock Secure data aggregation in wireless sensor network: a survey.
\newblock In {\em Proceedings of the 6th Australasian Information Security
  Conference (AISC)}, pages 93--105, Darlinghurst, Australia, January 2008.

\bibitem{rfc4948}
L.~Andersson, E.~Davies, and L.~Zhang.
\newblock {Report from the IAB workshop on Unwanted Traffic March 9-10, 2006}.
\newblock RFC 4948 (Informational), August 2007.

\bibitem{Azimi-Sadjadi2007}
B.~Azimi-Sadjadi, A.~Kiayias, A.~Mercado, and B.~Yener.
\newblock Robust key generation from signal envelopes in wireless networks.
\newblock In {\em Proceedings of the 2007 ACM Conference on Computer and
  Communications Security (CCS)}, pages 401--410, Alexandria, USA, October
  2007.

\bibitem{Becher2006}
A.~Becher, Z.~Benenson, and M.~Dornseif.
\newblock Tampering with motes: Real-world attacks on wireless sensor networks.
\newblock In {\em Sicherheit - Schutz und Zuverl{\"a}ssigkeit, Beitr{\"a}ge der
  3. Jahrestagung des Fachbereichs Sicherheit der Gesellschaft f{\"u}r
  Informatik e.v. (GI)}, pages 26--29, Magdeburg, Germany, February 2006.

\bibitem{Bo2007}
Su~Mon Bo, Hannan Xiao, Aderemi Adereti, James~A. Malcolm, and Bruce
  Christianson.
\newblock A performance comparison of wireless ad hoc network routing protocols
  under security attack.
\newblock In {\em Proceedings of the 3d International Symposium on Information
  Assurance and Security (IAS)}, pages 50--55, Manchester, United Kingdom,
  August 2007.

\bibitem{Castelluccia2005}
E.~Mykletun C.~Castelluccia, A. C-F.~Chan and G.~Tsudik.
\newblock Efficient and provably secure aggregation of encrypted data in
  wireless sensor networks.
\newblock {\em ACM Transactions Sensor Networks}, 5(3):1--36, 2009.

\bibitem{Chan2004}
H.~Chan, A.~Perrig, and D.~X. Song.
\newblock Random key predistribution schemes for sensor networks.
\newblock In {\em IEEE Symposium on Security and Privacy (S{\&}P )}, page 197,
  Berkeley, USA, May 2003.

\bibitem{Chan2006}
Haowen Chan, Adrian Perrig, and Dawn~Xiaodong Song.
\newblock Secure hierarchical in-network aggregation in sensor networks.
\newblock In {\em Proceedings of the 13th ACM Conference on Computer and
  Communications Security (CCS)}, pages 278--287, Alexandria, USA, October
  2006.

\bibitem{Chen2002}
D.~Chen, S.~Garg, and K.~S. Trivedi.
\newblock Network survivability performance evaluation: a quantitative approach
  with applications in wireless ad-hoc networks.
\newblock In {\em Proceedings of the 5th International Symposium on Modeling
  Analysis and Simulation of Wireless and Mobile Systems (MSWiM)}, pages
  61--68, Atlanta, USA, September 2002.

\bibitem{Chiang07}
J.~T. Chiang and Y-C. Hu.
\newblock Cross-layer jamming detection and mitigation in wireless broadcast
  networks.
\newblock In {\em Proceedings of the 13th Annual International Conference on
  Mobile Computing and Networking (MOBICOM)}, pages 346--349, Montr{\'e}al,
  Canada, September 2007.

\bibitem{Kim2006}
K.~M.~Shazzad D.~S.~Kim and J.~S. Park.
\newblock A framework of survivability model for wireless sensor network.
\newblock In {\em Proceedings of the The 1st International Conference on
  Availability, Reliability and Security (ARES)}, pages 515--522, Vienna,
  Austria, April 2006.

\bibitem{Girao2006}
J.~Girão D.~Westhoff and M.~Acharya.
\newblock Concealed data aggregation for reverse multicast traffic in sensor
  networks: Encryption, key distribution, and routing adaptation.
\newblock {\em IEEE Transactions on Mobile Computing}, 5(10):1417--1431, 2006.
\newblock Member-Westhoff, Dirk and Student Member-Girao, Joao and Student
  Member-Acharya, Mithun.

\bibitem{deswarte2006}
Y.~Deswarte and D.~Powell.
\newblock Internet security: an intusion tolerance approach.
\newblock In {\em Proceedings of the IEEE ISSN}, volume~2, pages 432--41, New
  York, USA, February 2006.

\bibitem{Djenouri2005}
D.~Djenouri, L.~Khelladi, and N.~Badache.
\newblock A survey of security issues in mobile ad hoc and sensor networks.
\newblock {\em IEEE Communications Surveys and Tutorials}, 7(1-4):2--28,
  October 2005.

\bibitem{domingo02}
J.~Domingo-Ferrer.
\newblock A provably secure additive and multiplicative privacy homomorphism.
\newblock In {\em 5th International Conference on Information Security (ISC)},
  pages 471--483, Sao Paulo, Brazil, September 2002.

\bibitem{Ellison1999}
R.~J. Ellison, R.~C. Linger, T.~Longstaff, and N.~R. Mead.
\newblock Survivable network system analysis: A case study.
\newblock {\em IEEE Software}, 16(4):70--77, July 1999.

\bibitem{Ganesan2001}
D.~Ganesan, R.~Govindan, S.~Shenker, and D.~Estrin.
\newblock Highly-resilient, energy-efficient multipath routing in wireless
  sensor networks.
\newblock In {\em Proceedings of the 2nd ACM Interational Symposium on Mobile
  Ad Hoc Networking and Computing (MobiHoc)}, pages 251--254, Long Beach, USA,
  October 2001.

\bibitem{Przydatek2007}
B.~Przydatek H.~Chan, A.~Perrig and D.~Song.
\newblock {SIA:} secure information aggregation in sensor networks.
\newblock {\em Journal of Computer Security}, 15(1):69--102, 2007.

\bibitem{Elyes2008}
E.~Hamida, G.~Chelius, and J-M. Gorce.
\newblock Scalable versus accurate physical layer modeling in wireless network
  simulations.
\newblock In {\em 22nd Workshop on Principles of Advanced and Distributed
  Simulation}, pages 127--134, Roma, Italy, June 2008.

\bibitem{HuEvans03}
L.~Hu and D.~Evans.
\newblock Secure aggregation for wireless network.
\newblock In {\em Symposium on Applications and the Internet Workshops
  (SAINT)}, pages 384--394, Orlando, USA, January 2003.

\bibitem{Hu2004}
L.~Hu and D.~Evans.
\newblock Using directional antennas to prevent wormhole attacks.
\newblock In {\em Proceedings of the Network and Distributed System Security
  Symposium (NDSS)}, pages 1--11, San Diego, USA, February 2004.

\bibitem{Hu2003a}
Y-C. Hu, A.~Perrig, and D.~B. Johnson.
\newblock Packet leashes: A defense against wormhole attacks in wireless
  networks.
\newblock In {\em 22nd Annual Joint Conference of the IEEE Computer and
  Communications Societies}, pages 1976--1986, San Fransisco, USA, April 2003.

\bibitem{Hu2005ariadne}
Y.~C. Hu, A.~Perrig, and D.~B. Johnson.
\newblock Ariadne: a secure on-demand routing protocol for ad hoc networks.
\newblock {\em Wireless Networks}, 11(1-2):21--38, January 2005.

\bibitem{Huang2004}
Y.~Huang and W.~Lee.
\newblock Attack analysis and detection for ad hoc routing protocols.
\newblock In {\em 7th International Symposium on Recent Advances in Intrusion
  Detection (RAID)}, pages 125--145, Sophia Antipolis, France, September 2004.

\bibitem{Dong2008}
B.~Bavar J.~Dong, K. E.~Ackermann and C.~Nita-Rotaru.
\newblock Mitigating attacks against virtual coordinate based routing in
  wireless sensor networks.
\newblock In {\em Proceedings of the 1st ACM Conference on Wireless Network
  Security (WISEC)}, pages 89--99, Alexandria, USA, March 2008.

\bibitem{Hur2005}
S-M.~Hong J.~Hur, Y.~Lee and H.~Yoon.
\newblock Trust management for resilient wireless sensor networks.
\newblock In {\em 8th International Conference on Information Security and
  Cryptology (ICISC)}, pages 56--68, Seoul, Korea, December 2005.

\bibitem{Mache2008}
C-Y.~Wan J.~Mache and M.~D. Yarvis.
\newblock Exploiting heterogeneity for sensor network security.
\newblock In {\em Proceedings of the 5th Annual IEEE Communications Society
  Conference on Sensor, Mesh and Ad Hoc Communications and Networks (SECON)},
  pages 591--593, San Francisco, USA, June 2008.

\bibitem{Jadia2004}
P.~Jadia and A.~Mathuria.
\newblock Efficient secure aggregation in sensor networks.
\newblock In {\em 11th International Conference on High Performance Computing
  (HiPC)}, pages 40--49, Bangalore, India, December 2004.

\bibitem{dsr96}
D.~B. Johnson and D.~A. Maltz.
\newblock Dynamic source routing in ad hoc wireless networks.
\newblock In Springer US, editor, {\em Mobile Computing}, volume 353, pages
  153--181, 1996.

\bibitem{Karlof2003}
C.~Karlof and D.~Wagner.
\newblock Secure routing in wireless sensor networks: attacks and
  countermeasures.
\newblock {\em Ad Hoc Networks}, 1(2-3):293--315, August 2003.

\bibitem{gpsr00}
B.~Karp and H.~T. Kung.
\newblock Gpsr: greedy perimeter stateless routing for wireless networks.
\newblock In {\em Proceedings of the 6th annual international conference on
  Mobile computing and networking}, pages 243--254, Boston, USA, August 2000.

\bibitem{Kim2005}
K-H. Kim, W-D. Jung, J-S. Park, H-G. Seo, S-H. Jo, C-M. Shin, S-M. Park, and
  H-N. Kim.
\newblock A resilient multipath routing protocol for wireless sensor networks.
\newblock In {\em 4th International Conference on Networking (ICN)}, pages
  1122--1129, Ile de la Reunion, April 2005.

\bibitem{LiYang2006}
X.~Li and D.~Yang.
\newblock A quantitative survivability evaluation model for wireless sensor
  networks.
\newblock In {\em IEEE International Conference on Networking, Sensing and
  Control}, pages 727--732, Okayama, Japan, March 2006.

\bibitem{limaSurvey09}
M.N. Lima, A.L. dos Santos, and G.~Pujolle.
\newblock A survey of survivability in mobile ad hoc networks.
\newblock {\em IEEE Communications Surveys and Tutorials}, 11(1):1--3, January
  2009.

\bibitem{lima2008}
M.N. Lima, H.W.da Silva, A.L. dos Santos, and G.Pujolle.
\newblock Survival multipath routing for manets.
\newblock In {\em Network Operations and Management Symposium}, pages 425--432,
  Salvador, Brazil, April 2008.

\bibitem{Conti2008}
L.~V.~Mancini M.~Conti, R. Di~Pietro and A.~Mei.
\newblock Emergent properties: detection of the node-capture attack in mobile
  wireless sensor networks.
\newblock In {\em Proceedings of the 1st ACM Conference on Wireless Network
  Security (WISEC)}, pages 214--219, Alexandria, USA, March 2008.

\bibitem{Mahimkar2004}
A.~Mahimkar and T.S. Rappaport.
\newblock Securedav: a secure data aggregation and verification protocol for
  sensor networks.
\newblock In {\em IEEE Global Telecommunications Conference (GLOBECOM)},
  volume~4, pages 2175--2179, Dallas, USA, November 2004.

\bibitem{gbr01}
Curt~Schurgers Mani and Mani~B. Srivastava.
\newblock Energy efficient routing in wireless sensor networks.
\newblock In {\em Military Communications Conference Proceedings on
  Communications for Network-Centric Operations: Creating the Information
  Force}, volume~1, pages 357--361, McLean, USA, October 2001.

\bibitem{Marti2000}
S.~Marti, T.~J. Giuli, K.~Lai, and M.~Baker.
\newblock Mitigating routing misbehavior in mobile ad hoc networks.
\newblock In {\em 6th Annual International Conference on Mobile Computing and
  Networking (MobiCom)}, pages 255--265, Boston, USA, August 2000.

\bibitem{Aguilar2009}
C.~A. Melchor, B.~Ait-Salem, P.~Gaborit, and K.~Tamine.
\newblock Active detection of node replication attacks.
\newblock {\em International Journal of Computer Science and Network Security
  (IJCSNS)}, 9:13--21, February 2009.

\bibitem{Michiardi2002}
P.~Michiardi and R.~Molva.
\newblock Core: a collaborative reputation mechanism to enforce node
  cooperation in mobile ad hoc networks.
\newblock In {\em 6th Joint Working Conference on Communications and Multimedia
  Security, Advanced Communications and Multimedia Security}, pages 107--121,
  Portoroz, Slovenia, September 2002.

\bibitem{Molva2003}
R.~Molva and P.~Michiardi.
\newblock Security in ad hoc networks.
\newblock In {\em 8th International Conference on Personal Wireless
  Communications (PWC)}, pages 756--775, Venice, Italy, September 2003.

\bibitem{Sastry2003}
U.~Shankar N.~Sastry and D.~Wagner.
\newblock Secure verification of location claims.
\newblock In {\em Proceedings of the 2003 ACM Workshop on Wireless Security
  (WiSe)}, pages 1--10, San Diego, USA, September 2003.

\bibitem{Newsome2004}
J.~Newsome, E.~Shi, D.~Song, and A.~Perrig.
\newblock The sybil attack in sensor networks: analysis {\&} defenses.
\newblock In {\em Proceedings of the Third International Symposium on
  Information Processing in Sensor Networks (IPSN)}, pages 259--268, Berkeley,
  USA, April 2004.

\bibitem{Ning2005}
P.~Ning and K.~Sun.
\newblock How to misuse aodv: a case study of insider attacks against mobile
  ad-hoc routing protocols.
\newblock {\em Ad Hoc Networks}, 3(6):795--819, 2005.

\bibitem{Haas02}
P.~Papadimitratos and Z.~Haas.
\newblock Secure rotuing for mobile ad hoc networks.
\newblock In {\em Communication Networks and Distributed Systems Modeling and
  Simulation Conference (CNDS)}, pages 27--31, San Antonio, Texas, January
  2002.

\bibitem{Parno2005}
B.~Parno, A.~Perrig, and V.D. Gligor.
\newblock Distributed detection of node replication attacks in sensor networks.
\newblock In {\em IEEE Symposium on Security and Privacy (S{\&}P)}, pages
  49--63, Oakland, USA, May 2005.

\bibitem{Perkins1994}
C.~E. Perkins and P.~Bhagwat.
\newblock Highly dynamic destination-sequenced distance-vector routing (dsdv)
  for mobile computers.
\newblock {\em ACM SIGCOMM Computer Communication Review}, 24(4):234--244,
  October 1994.

\bibitem{Perrig04secu}
A.~Perrig, J.~Stankovic, D.~Wagner, and C.~Rosenblatt.
\newblock Security in wireless sensor networks.
\newblock {\em Communications of the ACM}, 47(6):53--57, June 2004.

\bibitem{Perrig2001}
A.~Perrig, R.~Szewczyk, V.~Wen, D.~E. Culler, and J.~D. Tygar.
\newblock Spins: security protocols for sensor netowrks.
\newblock In {\em 7th Annual International Conference on Mobile Computing and
  Networks}, pages 189--199, Rome, Italy, July 2001.

\bibitem{Qian2007}
Y.~Qian, K.~Lu, and D.~Tipper.
\newblock Towards survivable and secure wireless sensor networks.
\newblock In {\em Proceedings of the 26th IEEE International Performance
  Computing and Communications Conference (IPCCC)}, pages 442--448, New
  Orleans, USA, April 2007.

\bibitem{Chen2006}
J.~M. Park M. T.~Refaei R.~Chen, M.~Snow and M.~Eltoweissy.
\newblock Defense against routing disruption attacks in mobile ad hoc networks.
\newblock In {\em Proceedings of the Global Telecommunications Conference
  (GLOBECOM)}, San Francisco, CA, USA, November 2006.

\bibitem{Rachedi2008}
A.~Rachedi.
\newblock {\em Contributions à la sécurité dans les réseaux mobiles Ad-hoc}.
\newblock PhD thesis, University of Avignon, France, November 2008.

\bibitem{Roosta2007}
T.~Roosta, S.~Pai, P.~Chen, S.~Sastry, and S.B. Wicker.
\newblock Inherent security of routing protocols in ad-hoc and sensor networks.
\newblock In {\em Proceedings of the Global Communications Conference
  (GLOBECOM)}, pages 1273--1278, Washington, USA, November 2007.

\bibitem{Dahill2002}
K.~Sanzgiri, B.Dahill, B.~N. Levine, C.~Shields, and E.~M. Belding-Royer.
\newblock A secure routing protocol for ad hoc networks.
\newblock In {\em 10th IEEE International Conference on Network Protocols
  (ICNP)}, pages 78--89, Paris, France, November 2002.

\bibitem{rwr02}
S.~D. Servetto and G.~Barrenechea.
\newblock Constrained random walks on random graphs: routing algorithms for
  large scale wireless sensor networks.
\newblock In {\em Proceedings of the 1st ACM International Workshop on Wireless
  Sensor Networks and Applications}, pages 12--21, Atlanta, USA, September
  2002.

\bibitem{Sterbenz2002}
J.~P.~G. Sterbenz, R.~Krishnan, R.~Hain, A.W. Jackson, D.~Levin, R.~Ramanathan,
  and J.~Zao.
\newblock Survivable mobile wireless networks: issues, challenges, and research
  directions.
\newblock In {\em Proceedings of the 2002 ACM Workshop on Wireless Security
  (WiSe)}, pages 31--40, Atlanta, USA, September 2002.

\bibitem{Tague2009a}
Patrick Tague, Mingyan Li, and Radha Poovendran.
\newblock Mitigation of control channel jamming under node capture attacks.
\newblock {\em IEEE Transactions on Mobile Computing}, 8(9):1221--1234, 2009.

\bibitem{Tague2008t}
Patrick Tague, Sidharth Nabar, James~A. Ritcey, David Slater, and Radha
  Poovendran.
\newblock Throughput optimization for multipath unicast routing under
  probabilistic jamming.
\newblock In {\em Proceedings of the IEEE 19th International Symposium on
  Personal, Indoor and Mobile Radio Communications (PIMRC)}, pages 1--5,
  Cannes, France, September 2008.

\bibitem{Tague2009}
Patrick Tague, David Slater, Jason Rogers, and Radha Poovendran.
\newblock Evaluating the vulnerability of network traffic using joint security
  and routing analysis.
\newblock {\em IEEE Transactions on Dependable and Secure Computing (TDSC)},
  6(2):111--123, 2009.

\bibitem{Toledo2008}
A.L. Toledo and Xiaodong Wang.
\newblock Robust detection of mac layer denial-of-service attacks in csma/ca
  wireless networks.
\newblock {\em IEEE Transactions on Information Forensics and Security},
  3(3):347--358, September 2008.

\bibitem{Tubaishat2004}
Malik Tubaishat, Jian Yin, Biswajit Panja, and Sanjay Madria.
\newblock A secure hierarchical model for sensor network.
\newblock {\em SIGMOD}, 33(1):7--13, 2004.

\bibitem{Du2003}
Y.~S.~Han W.~Du, J.~Deng and P.K. Varshney.
\newblock A witness-based approach for data fusion assurance in wireless sensor
  networks.
\newblock In {\em IEEE Global Telecommunications Conference (GLOBECOM)},
  volume~3, pages 1435--1439, San Francisco, USA, December 2003.

\bibitem{Wagner2004}
D.~Wagner.
\newblock Resilient aggregation in sensor networks.
\newblock In {\em Proceedings of the 2nd ACM Workshop on Security of ad hoc and
  Sensor Networks (SASN)}, pages 78--87, Washington, USA, October 2004.

\bibitem{Walters2007}
John~Paul Walters, Zhengqiang Liang, Weisong Shi, and Vipin Chaudhary.
\newblock Wireless sensor network security: A survey," in book chapter of
  security.
\newblock In {\em Security in Distributed, Grid, Mobile, and Pervasive
  Computing}, pages 0--849. CRC Press, 2007.

\bibitem{watteyne08energy}
Thomas Watteyne.
\newblock {\em Energy-Efficient Self-Organization for Wireless Sensor
  Networks}.
\newblock PhD thesis, INSA de Lyon, November 2008.
\newblock number 2008-ISAL-0082.

\bibitem{Wood2002}
A.D. Wood and J.A. Stankovic.
\newblock Denial of service in sensor networks.
\newblock {\em Computer}, 35(10):54--62, October 2002.

\bibitem{Wood03jam}
Anthony~D. Wood, John~A. Stankovic, and Sang~Hyuk Son.
\newblock Jam: A jammed-area mapping service for sensor networks.
\newblock In {\em Proceedings of the 24th IEEE Real-Time Systems Symposium
  (RTSS)}, pages 286--297, Cancun, Mexico, December 2003.

\bibitem{ChenMakki2009}
K.~Makki K.~Yen X.~Chen and N.~Pissinou.
\newblock Sensor network security: a survey.
\newblock {\em IEEE Communications Surveys \& Tutorials}, 11(2):52--73, 2009.

\bibitem{Chen2007}
K.~Yen X.~Chen, K.~Makki and N.~Pissinou.
\newblock Node compromise modeling and its applications in sensor networks.
\newblock In {\em Proceedings of the 12th IEEE Symposium on Computers and
  Communications (ISCC)}, pages 575--582, Aveiro, Portugal, July 2007.

\bibitem{Chen2008}
K.~Yen X.~Chen, K.~Makki and N.~Pissinou.
\newblock Attack distribution modeling and its applications in sensor network
  security.
\newblock {\em EURASIP Journal on Wireless Communications and Networking},
  2008:1--11, 2008.

\bibitem{Xu2006}
Wenyuan Xu, Ke~Ma, W.~Trappe, and Yanyong Zhang.
\newblock Jamming sensor networks: attack and defense strategies.
\newblock {\em IEEE Network}, 20(3):41--47, May 2006.

\bibitem{Xu2004}
Wenyuan Xu, Timothy Wood, Wade Trappe, and Yanyong Zhang.
\newblock Channel surfing and spatial retreats: defenses against wireless
  denial of service.
\newblock In {\em Proceedings of the 2004 ACM Workshop on Wireless Security
  (WiSe)}, pages 80--89, Philadelphia, USA, October 2004.

\bibitem{Hu2003}
A.~Perrig Y-C.~Hu and D.B. Johnson.
\newblock Rushing attacks and defense in wireless ad hoc network routing
  protocols.
\newblock In {\em Proceedings of the 2003 ACM Workshop on Wireless Security
  (WiSe)}, pages 30--40, San Diego, USA, September 2003.

\bibitem{Law2009}
L.~V. Hoesel J. Doumen P.~Hartel Y.~W.~Law, M.~Palaniswami and P.~Havinga.
\newblock Energy-efficient link-layer jamming attacks against wireless sensor
  network mac protocols.
\newblock {\em ACM Transactions on Sensor Networks}, 5(1):1--38, 2009.

\bibitem{Yang2008}
Yi~Yang, Xinran Wang, Sencun Zhu, and Guohong Cao.
\newblock Sdap: a secure hop-by-hop data aggregation protocol for sensor
  networks.
\newblock In {\em Proceedings of the 7th ACM Interational Symposium on Mobile
  Ad Hoc Networking and Computing (MobiHoc)}, pages 356--367, Florence, Italy,
  May 2006.

\bibitem{Ye2005}
Fan Ye, Haiyun Luo, Songwu Lu, and Lixia Zhang.
\newblock Statistical en-route filtering of injected false data in sensor
  networks.
\newblock {\em IEEE Journal on Selected Areas in Communications},
  23(4):839--850, 2005.

\bibitem{Yu2005}
Zhen Yu and Yong Guan.
\newblock A dynamic en-route scheme for filtering false data injection in
  wireless sensor networks.
\newblock In {\em 25th IEEE International Conference on Computer
  Communications, Joint Conference of the IEEE Computer and Communications
  Societies (INFOCOM)}, Barcelona, Spain, April 2006.

\bibitem{Zhen2003}
Jane Zhen and Sampelli Srinivas.
\newblock Preventing replay attacks for secure routing in ad hoc networks.
\newblock In {\em Proceedings of the Second International Conference on Ad-Hoc,
  Mobile, and Wireless Networks (ADHOC-NOW)}, pages 140--150, Montreal, Canada,
  October 2003.

\bibitem{Zhu2007}
Zhengjian Zhu, Qingping Tan, and Peidong Zhu.
\newblock An effective secure routing for false data injection attack in
  wireless sensor network.
\newblock In {\em 10th Asia-Pacific Network Operations and Management
  Symposium, Managing Next Generation Networks and Services (APNOMS)}, pages
  457--465, Sapporo, Japan, October 2007.

\bibitem{Znaidi2009}
W.~{Z}naidi and M.~{M}inier.
\newblock {P}roposition de gestion des cl{\'e}s et de contr{\^o}le d'acc{\`e}s
  dans un r{\'e}seau de capteurs.
\newblock In {\em 10{\`e}me {J}ourn{\'e}es {D}octorales {I}nformatique et
  {R}{\'e}seau}, pages 1--10, {B}elfort, {F}rance, February 2009.

\bibitem{Znaidi2008}
W.~Znaidi, M.~Minier, and J-P. Babau.
\newblock Detecting wormhole attacks in wireless networks using local
  neighborhood information.
\newblock In {\em Proceedings of the IEEE 19th International Symposium on
  Personal, Indoor and Mobile Radio Communications (PIMRC)}, pages 1--5,
  Cannes, France, September 2008.

\bibitem{Znaidi_Minier_Babau_08}
W.~Znaidi, M.~Minier, and J-P. Babau.
\newblock An ontology for attacks in wireless sensor networks.
\newblock Research Report RR-6704, INRIA, 2008.

\end{thebibliography}

\end{document}